\documentclass[format=acmsmall, review=false, screen=true]{acmart}

\usepackage{booktabs} 
\usepackage{fancyhdr}
\usepackage{subfig}
\usepackage{graphicx}
\usepackage{array}
\usepackage{url}
\usepackage{footmisc} 
\usepackage{fancybox}
\usepackage{multirow}
\usepackage{amssymb}
\usepackage{color}
\usepackage{colortbl}
\usepackage{fancyhdr}
\usepackage{subfig}
\usepackage{diagbox}
\usepackage{bbding}
\usepackage{placeins}
\usepackage{balance}
\usepackage{listings}
\usepackage{amsmath}
\usepackage{amsfonts}
\usepackage{stackrel}
\usepackage{svg}
\usepackage{setspace}
\usepackage{natbib}
\setcitestyle{square,comma,numbers}
\PassOptionsToPackage{prologue,dvipsnames}{xcolor}
\usepackage[dvipsnames]{xcolor}
\usepackage[most]{tcolorbox}
\usepackage[linesnumbered, ruled,vlined]{algorithm2e}
\usepackage{hyperref}
\setcopyright{acmcopyright}
\acmJournal{TOSEM}

\newcommand{\todo}[1]{}
\renewcommand{\todo}[1]{{\color{red} TODO: {#1}}}

\begin{document}
        \fancyfoot{} 
        \renewcommand{\headrulewidth}{0pt} 
        \fancypagestyle{titlepage}{
        \fancyhf{} 
        \renewcommand{\headrulewidth}{0pt} 
        }
	\title{When ChatGPT Meets Smart Contract Vulnerability Detection: How Far Are We?}
	
	\author{Chong Chen}
	\affiliation{%
		\institution{Sun Yat-sen University}
        \country{China}
	}
    \affiliation{%
		\institution{The State Key Laboratory of Blockchain and Data Security, Zhejiang University}
        \country{China}
	}
	\email{chench578@mail2.sysu.edu.cn}

        \author{Jianzhong Su}
	\affiliation{%
		\institution{Sun Yat-sen University}
            \country{China}
	}
	\email{sujzh3@mail2.sysu.edu.cn}

        \author{Jiachi Chen}
 \authornote{corresponding author}
	\affiliation{%
		\institution{Sun Yat-sen University}
        \country{China}
	}
    \affiliation{%
		\institution{The State Key Laboratory of Blockchain and Data Security, Zhejiang University}
        \country{China}
	}
	\email{chenjch86@mail.sysu.edu.cn}
	
	\author{Yanlin Wang}
	\affiliation{%
		\institution{Sun Yat-sen University}
            \country{China}
	}
	\email{wangylin36@mail.sysu.edu.cn}
	
	\author{Tingting Bi}
	\affiliation{%
		\institution{University of Western Australia}
            \country{Australia}
	}
	\email{Tingting.Bi@uwa.edu.au}

        \author{Jianxing Yu}
	\affiliation{%
		\institution{Sun Yat-sen University}
            \country{China}
	}
	\email{yujx26@mail.sysu.edu.cn}

        \author{Yanli Wang}
	\affiliation{%
		\institution{Sun Yat-sen University}
            \country{China}
	}
	\email{wangyli58@mail2.sysu.edu.cn}

        \author{Xingwei Lin}
	\affiliation{%
		\institution{Zhejiang University}
            \country{China}
	}
	\email{xwlin.roy@gmail.com}
 
	\author{Ting Chen}
	\affiliation{%
		\institution{University of Electronic Science and Technology of China}
            \country{China}
	}
	\email{brokendragon@uestc.edu.cn}

        \author{Zibin Zheng}
	\affiliation{%
		\institution{Sun Yat-sen University}
            \country{China}
	}
	\email{zhzibin@mail.sysu.edu.cn}
	

	\UseRawInputEncoding
\pdfoutput=1
\begin{abstract}
With the development of blockchain technology, smart contracts have become an important component of blockchain applications. Despite their crucial role, the development of smart contracts may introduce vulnerabilities and potentially lead to severe consequences, such as financial losses. Meanwhile, large language models, represented by ChatGPT, have gained great attentions, showcasing great capabilities in code analysis tasks. In this paper, we presented an empirical study to investigate the performance of ChatGPT in identifying smart contract vulnerabilities. Initially, we evaluated ChatGPT's effectiveness using a publicly available smart contract dataset. Our findings discover that while ChatGPT achieves a high recall rate, its precision in pinpointing smart contract vulnerabilities is limited. Furthermore, ChatGPT's performance varies when detecting different vulnerability types. We delved into the root causes for the false positives generated by ChatGPT, and categorized them into four groups. Second, by comparing ChatGPT with other state-of-the-art smart contract vulnerability detection tools, we found that ChatGPT's F-score is lower than others for 3 out of the 7 vulnerabilities. In the case of the remaining 4 vulnerabilities, ChatGPT exhibits a slight advantage over these tools. Finally, we analyzed the limitation of ChatGPT in smart contract vulnerability detection, revealing that the robustness of ChatGPT in this field needs to be improved from two aspects: its \textit{uncertainty} in answering questions; and the \textit{limited length} of the detected code. In general, our research provides insights into the strengths and weaknesses of employing large language models, specifically ChatGPT, for the detection of smart contract vulnerabilities.
\end{abstract}
	
	%
	%
	


	\maketitle
        \thispagestyle{titlepage}
	\UseRawInputEncoding
\pdfoutput=1
\section{Introduction}
\label{Introduction}
In recent years, with the rapid development of blockchain platforms (e.g., Ethereum), smart contracts have been widely adopted in various domains~\cite{zheng2020overview}. However, the security issues arising from smart contract vulnerabilities have also resulted in significant economic losses~\cite{8976179}. These security concerns not only directly harm users' funds, but also have adverse effects on the overall development of the blockchain ecosystem. Therefore, detecting vulnerabilities in smart contracts has become an important task in recent years. 

Meanwhile, the introduction of ChatGPT~\cite{openai2023gpt4}, a Large Language Model (LLM) developed by the OpenAI team~\cite{openai2022}, has made a significant impact in various fields. 
ChatGPT's potential in computer programming has attracted significant attention in recent research~\cite{biswas2023role}. 
Some recent studies have explored the applicability of LLM in vulnerability detection~\cite{shafiq2023use,li2023test}. While there has been considerable analysis of ChatGPT's code capabilities in previous work, research on smart contract vulnerability detection remains relatively limited. In addition to ChatGPT, common LLM models include LLaMA~\cite{touvron2023llama} and PaLM2~\cite{anil2023palm}. LLaMA is a collection of foundation language models ranging from 7B to 65B parameters. PaLM2 is a large language model that builds on Google's~\cite{google} legacy of breakthrough research in machine learning and responsible artificial intelligence. Considering that GPT~\cite{gpt-api} was among the first commercially available large language models and continues to be widely used, it was our choice for this research. 

In this work, we embarked on empirical research focused on smart contract vulnerability detection using ChatGPT.
To explore the capabilities of ChatGPT in analyzing, understanding, and detecting smart contract vulnerabilities, we conducted an empirical study on its effectiveness, limitations, and comparison with other state-of-the-art smart contract vulnerability detection tools. In our exploration of using the GPT model for smart contract vulnerability detection, we initiated the assessment by evaluating its effectiveness using the wildely-used smartbugs dataset~\cite{durieux2020empirical}. Subsequently, we conducted a comparative analysis of the detection performance of the GPT model against 14 other SOTA tools. Lastly, we examined the limitations associated with applying GPT in the realm of smart contract vulnerability detection. Figure~\ref{fig:overview} provides an overview of our research methodology. We focus on the following research questions:

\begin{figure*}[t]
    \centering
    \includegraphics[width=1\linewidth, bb=0 0 930 350]{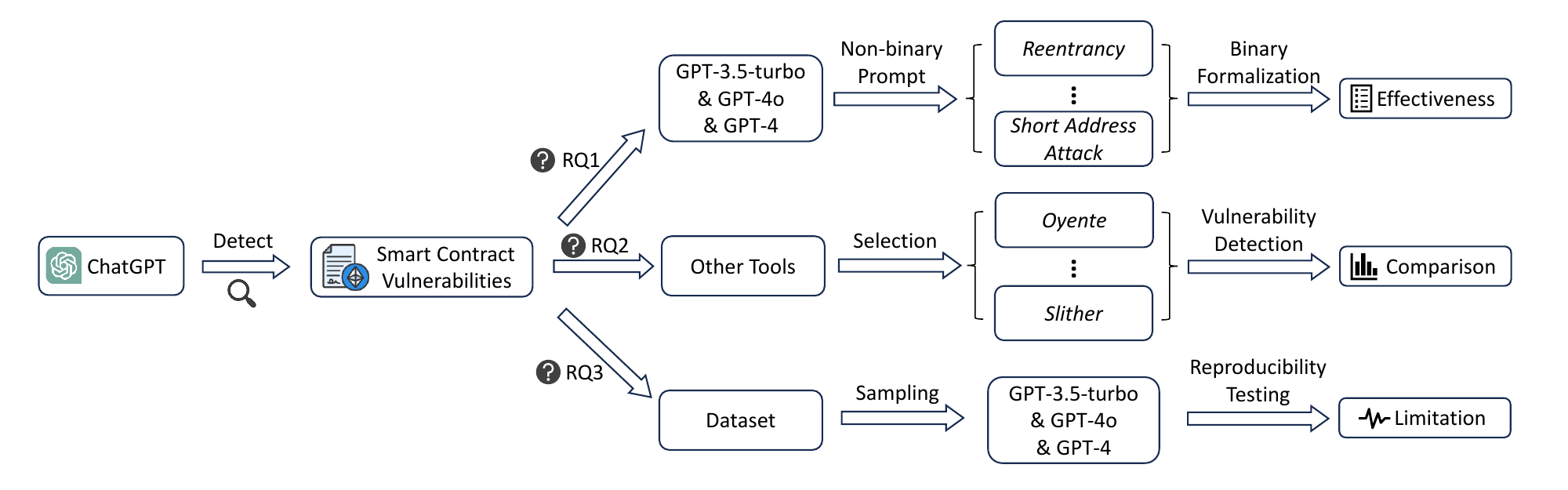}
    \caption{The overview of our research.}
    \label{fig:overview}
\end{figure*}

\noindent\textbf{\textit{RQ1. How effective is ChatGPT in detecting vulnerabilities in smart contracts?}}

We selected the widely-used smartbugs-curated dataset~\cite{durieux2020empirical}, which comprises 142 vulnerable smart contracts. The dataset contains nine categories of vulnerabilities as described in DASP10~\cite{dasp2018}. We designed a prompt for ChatGPT to detect these nine categories of vulnerability. The effectiveness of vulnerability detection is evaluated through various metrics (i.e., precision, recall, and F-score). Our findings indicated that while ChatGPT achieves a promising recall rate (i.e., 51.4\%, 83.6\% and 88.2\% for GPT-3.5, GPT-4o and GPT-4, respectively), its precision in pinpointing smart contract vulnerabilities is relatively low (19.7\%, 20.2\% and 22.6\% for GPT-3.5, GPT-4o and GPT-4, respectively). In addition, the detection performance of ChatGPT varies across different types of vulnerabilities. For example, GPT-3.5, GPT-4o and GPT-4 all perform well in detecting \textit{Unchecked Return Values}~\cite{gill2022finding}, however, all of them perform poorly in identifying \textit{Short Address Attack}~\cite{chen2023healthier}. Furthermore, we conducted a systematic analysis of false positives within ChatGPT, categorizing them into 4 distinct types: \textit{Protected Mechanism Bias}, \textit{Development Intent Bias}, \textit{Interfered by Comments} and \textit{Interference with Dead Code}.
\vspace{1ex}

\noindent\textbf{\textit{RQ2. How does ChatGPT perform compared to other tools?}}

In RQ1, we focus only on ChatGPT's effective in detecting certain vulnerabilities. To fully understand the level of capability of ChatGPT, it is important to compare its performance against other state-of-the-art tools. Through comparative analysis, we can elucidate the relative strengths and limitations of ChatGPT within the domain of vulnerability detection, thereby providing insights to inform future tool selection and development efforts. To evaluate ChatGPT's performance compared to other state-of-the-art smart contract vulnerability detection tools, we employed 14 vulnerability tools (e.g., Mythril~\cite{consensysmythril} and Confuzzius~\cite{confuzzius}) on the same smartbugs-curated~\cite{durieux2020empirical} dataset. 
Our results show that although ChatGPT outperformed other 14 tools in detecting specific vulnerabilities like \textit{Front Running} and \textit{denial of service}, the F1 scores are relatively low, measuring only 12.8\% and 15.9\%, respectively. For other vulnerability types, several tools demonstrate more effective detection capabilities than ChatGPT. However, in terms of detection speed, ChatGPT surpasses all other tools. Still, ChatGPT's detection failure rate surpasses that of 10 other tools, due to the token limitations of large language models.
\vspace{1ex}

\noindent\textbf{\textit{RQ3. What are the limitations of ChatGPT in detecting smart contract vulnerabilities?}}

As an artificial intelligence dialogue model, ChatGPT may produce different vulnerability detection results for the same smart contract when provided with the same prompt, potentially diminishing its reliability. 
To evaluate its reliability, we randomly sampled 50 smart contracts in the smartbugs-curated~\cite{durieux2020empirical} dataset and conducted vulnerability detection on them for five times with the same prompt. Our analysis showed that only 29 out of 50 contracts displayed entirely consistent results across all five round of detection. This suggests that 42\% of the evaluated smart contracts exhibit unstable detection outcomes. Additionally, the official documentation specifies the maximum token limit that the GPT model can handle. However, it does not provide precise information about limit of the associated code size, leaving developers without clear guidance on this aspect. As a result, we conducted experiments on both the gpt-3.5-turbo and gpt-4 models to approximate the maximum code size they can accommodate.

We summarized our main contributions as follows:
\begin{itemize}

\item We applied the gpt-3.5-turbo, the gpt-4o and the gpt-4 model to perform vulnerability detection on the smartbugs-curated dataset to evaluate their effectiveness. We provided evidence that ChatGPT has different detection performance for different vulnerabilities, with a relatively high recall rate but low precision. We also summarized 4 root causes of the false positives generated by ChatGPT.
\begin{spacing}{0.7}
\end{spacing}
\item We compared ChatGPT with 14 smart contract vulnerability detection tools in terms of effectiveness and efficiency. By conducting tests on 142 smart contracts, we identified two key insights into the strengths and limitations of ChatGPT. These insights represent valuable contributions to the ongoing research in leveraging large language models (LLM) for vulnerability detection.
\begin{spacing}{0.7}
\end{spacing}
\item We evaluatded stability of ChatGPT in detecting smart contract vulnerabilities, revealing that it is still uncertain in generating results of vulnerability detection. Additionally, we estimated the maximum code size that can be accommodated by both gpt-3.5-turbo and gpt-4.
\begin{spacing}{0.7}
\end{spacing}
\item We made all experimental data publicly (i.e.,   Zenodo repository: \url{https://zenodo.org/record/8332273}.), which could provide a guideline for developers and researchers to conduct the similar experiments for smart contact vulnerability detection. 

\end{itemize}

The organization of the rest of this paper is as follows. In Section~\ref{sec:background}, we provided background information on smart contracts and some common vulnerabilities in
them. In Sections ~\ref{sec:effectiveness}, ~\ref{sec:comparison}, and ~\ref{sec:limitations}, we elaborated on the experiments and conclusions of each RQ in detail. Specifically, in Section~\ref{sec:effectiveness}, we elaborated on the research on the effectiveness of ChatGPT in detecting smart contract vulnerabilities. In Section~\ref{sec:comparison}, we compared the performance of ChatGPT with multiple current smart contract vulnerability detection tools on the same dataset. In Section~\ref{sec:limitations}, we conducted research on the limitations of ChatGPT in detecting smart contract vulnerabilities. In Section~\ref{sec:discussion}, some aspects of this study are discussed. In Section~\ref{sec:related}, we presented some related work. In Section~\ref{sec:conclusion}, we summarized this study and provide future prospects.
	\UseRawInputEncoding
\pdfoutput=1
\section{Background}
\label{sec:background}
In this section, we briefly introduced background information about smart contracts and some common vulnerabilities. 

\subsection{Smart Contract}
Smart contracts are automated execution protocols based on blockchain technology, whose results can be widely recognized and agreed upon. The concept of smart contracts was initially introduced by Szabo in 1997~\cite{szabo1997formalizing}. This concept describes smart contracts as programs that facilitate digital verification without the need for trust and are self-executing. Accordingly, unlike traditional programs, smart contracts can be executed without the involvement of arbitrators or third-party institutions. Smart contracts are typically developed using high-level programming languages, such as Solidity~\cite{dannen2017introducing}, the most popular language for programming Ethereum smart contracts.

\subsection{Smart Contract Vulnerabilities} \label{sec:bkg_vul}
Smart contract vulnerabilities refer to security flaws or loopholes present in the smart contract, which can be exploited to manipulate contract rules or steal assets~\cite{9143290}. Due to the immubility of blockchian, once the smart contract is deployed on the blockchain, the code cannot be changed anymore, even the vulnerability is detected. Therefore, it may be maliciously exploited or cause irreparable economic losses~\cite{8976179}.

DASP~\cite{dasp2018} (Decentralized Application Security Project) is an initiative of the National Computing Centre (NCC) Group. It is an open and collaborative project to join efforts to discover smart contract vulnerabilities within the security community. DASP10 introduces 10 common vulnerabilities in smart contracts, such as reentrancy, access control, etc. Table~\ref{tab:types_of_vulnerabilities} lists the top 9 vulnerabilities in this list. The 10th type of vulnerability, known as Unknown Unknowns, refers to other vulnerabilities that may occur during iterative updates of smart contracts and will not be discussed further here.

Smartbugs-curated dataset~\cite{durieux2020empirical} is based on the DASP taxonomy. It provides a collection of vulnerable Solidity smart contracts organized according to the DASP taxonomy, as shown in Table~\ref{tab:types_of_vulnerabilities}. The first column of the table is the vulnerability name, the second column is the corresponding description, and the third column is the severity level. Specifically, Smartbugs-curated is a dataset provided by the Smartbugs project team containing 142 smart contracts, each contract code labeled with the specified vulnerability type. These contracts have certain representativeness and importance and can be used to test and verify the effectiveness of vulnerability detection tools. There are multiple sources of contracts in the dataset, including Ethereum and other public blockchain applications. Our research is based on the DASP10 and smartbugs-curated datasets.

\begin{table*}[t]
    \centering
    \small
        \caption{Top 9 smart contract vulnerabilities in DASP10 and their corresponding descriptions and levels(``Solidity'' level vulnerabilities are issues in how the contract is coded. ``EVM'' level ones come from the EVM execution model. ``Blockchain'' level ones are inherent to public blockchains.)}
        \resizebox{\linewidth}{!}{
            \begin{tabular}{l|l|l}
                \hline
                \textbf{Vulnerability} & \textbf{Description} & \textbf{Level}\\
                \hline
Reentrancy (RE) & Reentrant function calls make a contract to behave in an unexpected way & Solidity \\
Access Control (AC) & Failure to use function modifiers or use of tx.origin & Solidity \\
Arithmetic Issues (AI) & Integer over/underflows & Solidity \\
Unchecked Return Values (UR) & call(), callcode(), delegatecall() or send() fails and it is not checked & Solidity \\
Denial Of Service (DoS) & The contract is overwhelmed with time-consuming computations & Solidity \\
Bad Randomness (BR) & Malicious miner biases the outcome & Blockchain \\
Front Running (FR) & Two dependent transactions that invoke the same contract are included in one block & Blockchain \\
Time Manipulation (TM) & The timestamp of the block is manipulated by the miner & Blockchain \\
Short Address Attack (SAA) & EVM itself accepts incorrectly padded arguments & EVM \\
\hline
        \end{tabular}
        }
        \label{tab:types_of_vulnerabilities}
\end{table*}

\subsection{ChatGPT}
\label{section:chatgpt_bkg}
Large Language Models (LLMs) represent a pivotal advancement in natural language processing (NLP), revolutionizing various tasks like text generation, question answering, and dialogue systems. Among these, ChatGPT~\cite{openai2023gpt4}, developed by OpenAI~\cite{openai2022}, stands out as a leading LLM, building upon the successes of the GPT series, notably GPT-3.5 and GPT-4~\cite{openai2023gpt4}. While numerous LLMs continue to emerge, including ChatGLM~\cite{du2022glm}, Xunfei-Xinghuo~\cite{xunfei}, and ChatYuan-Large~\cite{chatyuan}, etc., studies indicate that ChatGPT consistently outperforms these models in terms of performance across most tasks~\cite{zhang2024llmeval}.

ChatGPT~\cite{openai2023gpt4} is a large language model developed by OpenAI~\cite{openai2022} company, which is an upgraded product of the GPT~\cite{gpt-api} series of models. ChatGPT uses deep learning structures such as transformers~\cite{LIN2022111} to pre-train and learn the statistical laws of language on large-scale text datasets. On this basis, perform task fine-tuning to gain powerful language processing capabilities such as text generation, question answering, and dialogue. During the training process, ChatGPT collected a large amount of text data from the Internet. The core principle of ChatGPT is the use of the Transformer~\cite{LIN2022111} architecture, which is an advanced neural network structure that enables the model to understand context, capture semantic correlations, and exhibit impressive intelligence when answering questions or generating text. The emergence of ChatGPT marks further progress in language understanding and the generation of large models. Currently, there are two widely used models for the GPT series: GPT-3.5 and GPT-4.
\begin{spacing}{1.5}
\end{spacing}
\noindent \textbf{GPT-3.5.} GPT-3.5 is an upgraded version of GPT-3 with fewer parameters that includes a fine-tuning process for machine learning algorithms~\cite{gpt-3-vs-gpt-3-5}. The fine-tuning process involves Reinforcement Learning~\cite{kaelbling1996reinforcement} with human feedback, which helps to improve the accuracy and effectiveness of the algorithms. The most capable and cost-effective model in the GPT-3.5 family is gpt-3.5-turbo which has been optimized for chat, but also works well for traditional completion tasks~\cite{gpt-3.5}.
\begin{spacing}{1.5}
\end{spacing}
\noindent \textbf{GPT-4.} Like previous GPT models, the GPT-4 base model was trained to predict the next word in a document and was trained using publicly available data (such as internet data) and data the OpenAI team has licensed~\cite{research-gpt4}. Moreover, it is a large-scale multimodal model that accepts input of image and text and outputs text~\cite{openai2023gpt4}. The GPT-4 model has a larger model size (1.76 trillion parameters~\cite{research-gpt4}) and better architecture, which can help it better understand the code, thus identifying potential problems more accurately. GPT-4o~\cite{gpt-4o} is an optimized version of GPT-4, designed to be more efficient in terms of computational resources while maintaining a similar or slightly reduced performance level. It’s often used in contexts where the full power of GPT-4 is not necessary, allowing for faster inference times and reduced costs.

This study only focused on the \textit{gpt-3.5-turbo}~\cite{gpt-3.5} , gpt-4o~\cite{gpt-4o} and \textit{gpt-4}~\cite{gpt-4} models mentioned above; nonetheless, updated versions of the models will emerge. Even with model updates, there exists a degree of stability. While the performance of models may undergo slight changes, some insights and discoveries remain significant. Model updates often enhance their overall understanding of large-scale corpora but may not necessarily yield substantial performance improvements in specific tasks. Therefore, our research results retain their reference value. Additionally, our findings also offer insights for other models, such as \textit{Claude}~\cite{claude} and \textit{Llama}~\cite{touvron2023llama}, as they share training principles with the models we utilized.


	\UseRawInputEncoding
\pdfoutput=1
\section{(RQ1) on Effectiveness of ChatGPT in Detecting Vulnerabilities}
\label{sec:effectiveness}

In this section, we explore the effectiveness of ChatGPT in detecting smart contract vulnerabilities with various metrics(i.e., precision, recall, and F1 score). We evaluate the performance of ChatGPT based on the top 9 smart contract vulnerabilities from the DASP10.
\subsection{Dataset}
We use the smartbugs-curated~\cite{durieux2020empirical} for our analysis. This dataset was chosen for the following key reasons: 1) its prevalent use within the smart contract vulnerability detection community~\cite{ferreira2020smartbugs,durieux2020empirical}; 2) each contract is meticulously labeled with associated vulnerabilities, ensuring the high-quality of data; 3) the straightforward nature of the dataset makes it suitable for the initial assessment of ChatGPT's vulnerability detection capabilities. Specifically, the average line of code of the contracts in the dataset is about 101.5 rows. Smartbugs-curated provides a collection of vulnerable Solidity smart contracts organized according to the DASP taxonomy, as shown in Table~\ref{tab:types_of_vulnerabilities}. For more information on the types of vulnerability detected and dataset in this experiment, refer to Section ~\ref{sec:bkg_vul}. It is important to note that all contracts in the dataset are flagged with vulnerabilities (positive), albeit with different types present in each. Our experiment focuses on specific vulnerability types within the contracts (e.g., when detecting \textit{reentrancy} vulnerability, contracts lacking such vulnerability are considered negative). Therefore, during testing, we evaluate the entire dataset as a unit. Contracts featuring a particular vulnerability type are labeled as positive, while those devoid of said vulnerability are labeled as negative.

\subsection{Experiment}
In order to evaluate the effectiveness of ChatGPT in detecting smart contract vulnerabilities, we use the GPT model API~\cite{gpt-api} to establish dialogue with the model. The models used in our experiment are the gpt-3.5-turbo-0613 model~\cite{gpt-3.5}, the gpt-4o-2024-05-13 model~\cite{gpt-4o} and the gpt-4-0613 model~\cite{gpt-4}. They are all the latest models in their respective series by the time of writing the paper. The parameter of $temperature$ was configured at 0.7 to strike a delicate equilibrium between fostering innovation and ensuring stability during text generation. All other parameters were maintained at their default values (e.g., $Top P$ set to 1 and $Frequency Penalty$ set to 0). Table~\ref{tab:Introduction_of_models} provides information on the three models. 

\begin{figure}[ht]
    \centering
    \includegraphics[width=280pt]{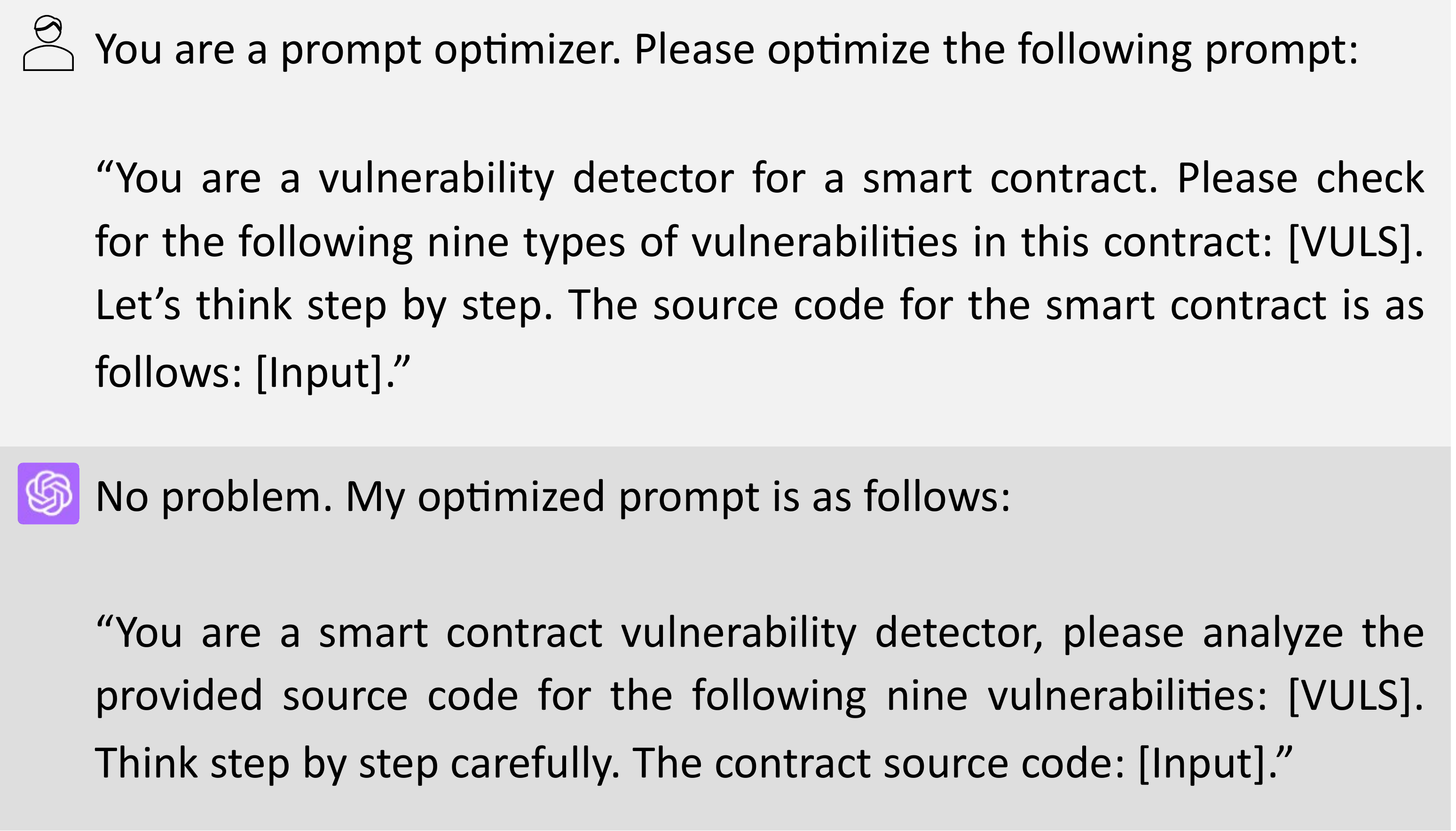}
    \caption{An example of prompt optimization using the GPT-4 model.}
    \label{fig:prompt_optimization}
\end{figure}

Models based on GPT architecture rely on a prompt-based learning approach~\cite{liu2022pretrain}. The effectiveness of the prompts significantly affects the overall performance of the models. To enhance the quality of the prompts, we employ ChatGPT to optimize our initial prompts and subsequently consolidate the generated prompts. In our study, we instruct ChatGPT to be a prompt optimizer with the task of optimizing the given prompt. This process helps us to generate refined prompts. We use the GPT model to generate multiple draft prompts, which undergo manual evaluation using sampled data to discern variances and derive high-quality prompts. Figure~\ref{fig:prompt_optimization} provides an illustration of prompt optimization using the GPT-4 model. In the context of smart contract vulnerability detection, we initiate the process by composing an initial task description. This draft prompt is then fed into the GPT-4 model for optimization, resulting in an optimized version of the draft. We sampled 20 smart contracts from the smartbugs-curated dataset~\cite{durieux2020empirical}, and tried to use different versions of the GPT-optimized prompts to instruct itself to perform vulnerability detection tasks.  In the end, the optimal prompt template was distilled as follows:
\begin{flushleft}
\texttt{You are [ROLE]. [TASK DESCRIPTION]. Think step by step, carefully. The input is [INPUT].}
\end{flushleft}
where the placeholder [ROLE] denotes the specific role assigned to ChatGPT. In this study, two main roles were designated: \textit{smart contract vulnerability detector} and \textit{text semantic analyzer}. \textit{smart contract vulnerability detector} is used to detect smart contract vulnerability tasks. The function of \textit{text semantic analyzer} is to perform semantic analysis on the conclusion of GPT detecting smart contract vulnerabilities, and achieve binary output of the detection conclusion (i.e., use ``1'' to indicate the existence of the vulnerability, and ``0'' to indicate the absence). We have created a new session to use this role, with the aim of avoiding the impact of context on the model. [TASK DESCRIPTION] describes the expected task that ChatGPT will perform. [INPUT] serves as a placeholder for the code under analysis.

We used the template above and provide the corresponding contract source code and command model to perform vulnerability detection tasks. In addition, in the prompt, we limited the scope of vulnerabilities to be detected for comparison with the ground truth in the dataset. The prompt template used is shown as follows:
\begin{flushleft}
\texttt{You are \textbf{a vulnerability detector for a smart contract. Here are nine common vulnerabilities. [VULS]. Check the following smart contract for the above vulnerabilities.} Think step by step, carefully. The input is [INPUT].}
\end{flushleft}
where \texttt{[VULS]} represents an introduction to the 9 vulnerabilities that need to be detected in this study. The introduction provides the names of each vulnerability along with their aliases, and this information is sourced from the official DASP10 website~\cite{dasp2018}. [INPUT] represents the source code of the smart contract to be detected.

\renewcommand{\arraystretch}{1.35}
\begin{table}[ht]
\rowcolors{1}{gray!15}{white}
    \centering
        \caption{Introduction to the two models used in the experiment.}
        
            \begin{tabular}{l|p{5cm}|l|l}
                \hline
                \textbf{Model} & \textbf{Description} & \textbf{Max Tokens} & \textbf{Training Data}\\
                \hline
gpt-3.5-turbo-0613 & Snapshot of gpt-3.5-turbo from June 13th 2023 with function calling data. & 4,096 tokens & Up to Sep 2021\\
\hline
gpt-4o-2024-05-13 & High-intelligence flagship model for complex, multi-step tasks. GPT-4o is cheaper and faster than GPT-4 Turbo. & 4,096 tokens & Up to Oct 2023\\
\hline
gpt-4-0613 & Snapshot of gpt-4 from June 13th 2023 with function calling data. & 8,192 tokens & Up to Sep 2021\\
\hline
        \end{tabular}
        
        \label{tab:Introduction_of_models}
\end{table}

Without specific constraints, ChatGPT can produce responses in a non-standardized format, which poses challenges for extracting coherent and meaningful insights. To enable a more systematic analysis of vulnerability detection results, one effective approach is to include commands in the prompt to limit the output format. However, imposing such format restrictions directly within prompts may also affect the performance of vulnerability detection in large language models. To explore whether restricting the format of the ChatGPT output would affect its vulnerability detection performance, we conducted an experiment on the entire dataset using the gpt-3.5-turbo model, while limiting the output format. 

Under the condition of limited output, the experimental result of gpt-3.5-turbo is shown in Table~\ref{tab:results_of_detection_restricted}. From the comparison between Table~\ref{tab:results_of_detection_restricted} and the left part of Table~\ref{tab:comparison_of_results}, it can be seen that, except for the vulnerability of \textit{Arithmetic Issues}, the limitation of the output format has a negative impact on the vulnerability detection performance of ChatGPT in all other vulnerabilities.

\renewcommand{\arraystretch}{0.9}
\begin{table}[ht]
    \centering
        \caption{Evaluation of gpt-3.5-turbo detection results on the dataset (restricted output).}
        
            \begin{tabular}{l|l|l|l}
                \hline
                \textbf{Vulnerability} & \textbf{Precision} & \textbf{Recall} & \textbf{F-measure} \\
                \hline
                Reentrancy & 26.0\% & 38.0\% & 31.5\% \\
                Access Control & 12.5\% & 65.0\% & 21.0\% \\
                Arithmetic Issues & 92.0\% & 60.0\% & 72.5\% \\
                Unchecked Return Values & 75.0\% & 25.0\% & 37.5\% \\
                Denial of Service & 9.5\% & 80.0\% & 17.0\% \\
                Bad Randomness & 95.0\% & 20.0\% & 33.3\% \\
                Front Running & 2.0\% & 15.0\% & 3.6\% \\
                Time manipulation & 8.0\% & 70.0\% & 14.5\% \\
                Short Address Attack & 0.0\% & 0.0\% & 0.0\% \\
                \hline
        \end{tabular}
        
        \label{tab:results_of_detection_restricted}
\end{table}
\renewcommand{\arraystretch}{1}

\begin{center}
\begin{tcolorbox}[colback=gray!10,
                  colframe=black,
                  width=\textwidth,
                  arc=1mm, auto outer arc,
                  boxrule=0.5pt,
                 ]
\noindent
\emph{\textbf{Finding 1:} Across the spectrum of eight vulnerability types, the vulnerability detection performance of ChatGPT may decrease due to output format limitations, indicating that limitations on ChatGPT output format may affect its ability.}
\end{tcolorbox}
\end{center}

Based on Finding 1, we attempted to use the inherent capabilities of large language models to refine the output. Specifically, once we obtain the vulnerability detection results from a large language model, we initiated a new conversation, feeding those results back into the model for semantic analysis. This ensures that the vulnerability detection output adheres to a unified format of ``[VUL]: 0/1''. The prompt template for this round of conversation is as follows:

\begin{flushleft}
\texttt{You are \textbf{a semantic analyzer of text. Here are nine common vulnerabilities. [VULS]. The following text is a vulnerability detection result for a smart contract. Use 0 or 1 to indicate whether there are specific types of vulnerabilities. For example: ``Reentrancy: 1''.} Think step by step, carefully. The input is [INPUT].}
\end{flushleft}

The effectiveness of the semantic analysis process requires validation. For example, if the GPT model identifies a contract as having both \textit{Reentrancy} and \textit{Arithmetic Issues} vulnerabilities, but the semantic analysis outcome indicates ``\textit{Reentrancy}: 1, \textit{Arithmetic Issues}: 0'', this renders the semantic analysis ineffective. To verify the effectiveness of the semantic analysis of the GPT model, we manually checked its standardized output. Out of all the 142 smart contracts, 50 were sampled from the dataset for evaluation. After normalization by the GPT model, 47 results were completely consistent with the original results. Two results show omissions, meaning that the normalized output did not include all the vulnerabilities in the original results. One result shows hallucinations, meaning that the original results did not indicate the existence of the vulnerability, while the normalized output indicated the existence of the vulnerability. The overall performance is shown in Table~\ref{tab:results_of_semantic}. This result indicates that the GPT model offers great reliability in semantic analyzing textual results and in standardizing output formats.

\renewcommand{\arraystretch}{1.25}
\begin{table}[ht]
\rowcolors{1}{gray!15}{white}
    \centering
        \caption{Result of semantic analysis and standardizing output formats of the GPT model on 50 sampled data.}
        
            \begin{tabular}{l|l|p{9cm}}
                \hline
                \textbf{Result} & \textbf{Number} & \textbf{Description}\\
                \hline
Correct & 47 & Completely consistent \\
\hline
Omissions & 2 & Missing some vulnerabilities inference included in original conclusions \\
\hline
Hallucination & 1 & Generating some vulnerabilities inference not included in original conclusion \\
\hline
        \end{tabular}
        
        \label{tab:results_of_semantic}
\end{table}
\renewcommand{\arraystretch}{1}

\subsection{Evaluation}
We used nine metrics to evaluate the detection results of ChatGPT, namely true positive (TP), true negative (TN), false positive (FP), false negative (FN), precision, recall, specificity, F1 score and standard deviation. TP and TN represent smart contracts with certain vulnerabilities correctly detected by ChatGPT and smart contracts without certain vulnerabilities, respectively. FP and FN indicate that ChatGPT has incorrectly detected smart contracts with or without certain vulnerabilities. Precision, recall, specificity, standard deviation and F1 score can be calculated as follows:

\begin{equation}\label{eq1}\rm
 Precision=\frac{\#TP}{\#TP+\#FP}
\end{equation}

\begin{equation}\label{eq2}\rm
 Recall=\frac{\#TP}{\#TP+\#FN}
\end{equation}

\begin{equation}\label{eq4}\rm
 Specificity=\frac{\#TN}{\#TN+\#FP}
\end{equation}

\begin{equation}\label{eq3}\rm
 F1=\frac{2*Precision*Recall}{Precision+Recall}
\end{equation}

\begin{equation}\label{eq5}\rm
 Std.=\sqrt{\frac{\sum_{i=1}^{5} (F1_i-\bar{F1})^2}{5}}
\end{equation}

    

\vspace{0.2cm}
The effectiveness of gpt-3.5-turbo in detecting vulnerabilities on the dataset is shown in the left part of Table~\ref{tab:comparison_of_results}. We repeated the same experiments on the smartbug dataset for five rounds to reduce the impact of the instability of the GPT model.

We replicated the same test by using the model  gpt-4o and gpt-4-0613. The middle and right part of Table~\ref{tab:comparison} shows the results of these experiments respectively. It can be seen that gpt-4o and gpt-4 has a significant improvement in the average recall rate of vulnerability detection. However, the precision has not changed much compared to gpt-3.5, leading to only a slight improvement of the overall F1 score.

Furthermore, it is worth noting that while using the gpt-3.5-turbo model, five smart contracts (3.5\%) were failed to be analyzed. Similarly, during the experimental phase involving the gpt-4 model, two smart contracts (1.4\%) encouter detection failures. These instances of detection failure can be attributed to ChatGPT's token limitations concerning input text. The gpt-3.5-turbo model has a maximum token limit of 4,096, while the gpt-4-0613 model can accommodate up to 8,192 tokens~\cite{gpt-api}. The length of smart contract codes for these unsuccessful detections exceeded the token threshold.

\begin{center}
\begin{tcolorbox}[colback=gray!10,
                  colframe=black,
                  width=\textwidth,
                  arc=1mm, auto outer arc,
                  boxrule=0.5pt,
                 ]
\noindent
\emph{\textbf{Finding 2:} GPT models may encounter detection failures in long smart contract code due to inherent token limitations.}
\end{tcolorbox}
\end{center}

\begin{table*}[ht]
    \centering
    \caption{Comparison of gpt-3.5-turbo and gpt-4 detection results on the dataset (The values in table represent percentages, and the column ``Std.'' represents the standard deviation of the F1 score indicators for five experiments.)}
    \resizebox{\linewidth}{!}{
        \begin{tabular}{l|l|l|l|l|l|l|l|l|l|l|l|l|l|l|l}
            \hline
            \multirow{2}{*}{\textbf{Vul.}} & \multicolumn{5}{c|}{\textbf{gpt-3.5-turbo}} & \multicolumn{5}{c|}{\textbf{gpt-4o}} & \multicolumn{5}{c}{\textbf{gpt-4}}\\
            \cline{2-16}
            & \textbf{Pre.} & \textbf{Re.} & \textbf{Spec.} & \textbf{F1} & \textbf{Std.} & \textbf{Pre.} & \textbf{Re.} & \textbf{Spec.} & \textbf{F1} & \textbf{Std.} & \textbf{Pre.} & \textbf{Re.} & \textbf{Spec.} & \textbf{F1} & \textbf{Std.}\\
            \hline
            RE & 28.3 & 41.9 & 77.5 & 33.8 & 0.0484 & 30.0 & 96.6 & 43.6 & 45.8 & 0.0224 & 33.5 & 96.8 & 41.4 & 49.8 & 0.0271\\
            AC & 14.3 & 72.2 & 20.2 & 23.9 & 0.0191 & 14.4 & 79.9 & 40.9 & 24.4 & 0.0237 & 16.7 & 95.6 & 29.8 & 28.4 & 0.0437\\
            AI & 25.8 & 53.3 & 77.2 & 34.8 & 0.0311 & 18.0 & 91.5 & 61.6 & 30.0 & 0.0231 & 22.0 & 96.0 & 63.0 & 35.7 & 0.0294\\
            UR & 58.5 & 59.6 & 73.3 & 59.0 & 0.0715 & 59.0 & 98.1 & 69.4 & 73.6 & 0.0338 & 53.6 & 92.7 & 67.6 & 67.9 & 0.0459\\
            DoS & 11.9 & 83.3 & 64.0 & 20.8 & 0.0195 & 4.8 & 96.0 & 36.6 & 9.1 & 0.0121 & 6.8 & 96.7 & 41.9 & 12.8 & 0.0168\\
            BR & 21.9 & 87.5 & 75.4 & 35.0 & 0.0800 & 38.0 & 100.0 & 91.9 & 54.8 & 0.0622 & 40.4 & 98.0 & 91.0 & 57.0 & 0.0418\\
            FR & 4.5 & 25.0 & 76.8 & 7.7 & 0.0000 & 4.9 & 70.0 & 63.9 & 9.1 & 0.0358 & 9.0 & 70.0 & 81.2 & 15.9 & 0.0080\\
            TM & 12.5 & 40.0 & 85.4 & 19.0 & 0.0088 & 12.1 & 100.0 & 76.6 & 21.5 & 0.0296 & 17.0 & 88.0 & 83.2 & 28.4 & 0.0447\\
            SAA & 0.0 & 0.0 & 78.7 & 0.0 & 0.0000 & 0.6 & 20.0 & 80.4 & 1.1 & 0.0256 & 4.5 & 60.0 & 90.8 & 8.4 & 0.0000\\
            \hline
            \textit{Avg.} & 19.7 & 51.4 & 69.8 & 26.0 & 0.0309 & 20.2 & 83.6 & 62.8 & 29.9 & 0.0298 & 22.6 & 88.2 & 65.5 & 33.8 & 0.0286\\
            \hline
        \end{tabular}
        }
    \label{tab:comparison_of_results}
\end{table*}











Table~\ref{tab:comparison_of_results} shows that ChatGPT achieves the best performance in detecting the ``\textit{Unchecked Return Values}'' vulnerability, with an F-score of 0.59, 0.74 and 0.68, respectively. Except for \textit{Front Running} and \textit{Short Address Attack}, gpt-4 achieves a good recall rate of more than 80\% for each type of vulnerability. However, the precision of both models is around 22\%, which indicates that ChatGPT has a large number of false positives during vulnerability detection. 

\begin{center}
\begin{tcolorbox}[colback=gray!10,
                  colframe=black,
                  width=\textwidth,
                  arc=1mm, auto outer arc,
                  boxrule=0.5pt,
                 ]
\noindent
\emph{\textbf{Finding 3:} ChatGPT displays distinct detection performances across various vulnerabilities. Of the 9 vulnerabilities examined, ChatGPT shows its strongest detection performance in detecting ``Unchecked Return Values" vulnerability.}
\end{tcolorbox}
\end{center}

\subsection{False Positive Analysis}
\label{subsec:fp_analysis}
In order to clarify the reasons for the false positives in ChatGPT, we manually analyzed each false positive case and classified the reasons for the false positives as follows:

\textbf{Protected Mechanism Bias.} We found that most false positives generated by ChatGPT stemmed from its excessive attention to the protected mechanisms embedded within the code. In smart contracts, some specific types of vulnerability have related protected mechanisms. For example, to prevent the occurrence of the ``Access Control'' vulnerability, developers often add relevant modifiers in their code, such as ``onlyOwner'', which strictly limits objects that can call specific functions, effectively preventing access control issues. 

\definecolor{darkgreen}{RGB}{100,150,100}
\begin{lstlisting}[caption=Example of using the onlyowner modifier in smart contracts.,label=lst:onlyowner]
pragma solidity ^0.4.15;
contract Unprotected{
    address private owner;
    modifier onlyowner {
        require(msg.sender==owner);
        _;
    }
    function Unprotected() public{
        owner = msg.sender;
    }
    function Protected(address _new) onlyowner{
        owner = _new;
    }
}
\end{lstlisting}

Listing~\ref{lst:onlyowner} is an example that uses the ``onlyowner'' modifier. This contract requires a function that is used to change the owner of this contract, and this function can only be called by the current contract owner. Line 4 is the ``onlyowner'' modifier. The function to change the contract owner in line 8 is an incorrect example because it is a public function that anyone can call. Line 11 is a correct example, which adds the ``onlyowner'' modifier to ensure that only the owner of the contract can call. Moreover, to prevent \textit{Arithmetic Issues} vulnerability, developers often add the \textit{SafeMath}~\cite{safemath} library to protect their code. As shown in Listing~\ref{lst:code1}, line 7 is an example of a developer calling the \textit{SafeMath} library to perform arithmetic operations, which can prevent integer overflow vulnerabilities.

However, ChatGPT overemphasizes these protected mechanisms. It might flag vulnerabilities (generate false positives) in contracts where such protected mechanisms are unnecessary. For example, if a smart contract devoid of arithmetic operations, it inherently avoids integer overflow vulnerabilities. Nevertheless, ChatGPT may incorrectly identify an integer overflow vulnerability in such cases, due to its assumption that the smart contract lacks the SafeMath library. Likewise, certain smart contracts, such as Listing~\ref{lst:code1}, might be designed solely for deposit or withdrawal actions, with no owner-specific functions. Such smart contracts do not necessitate modifiers like ``onlyOwner'' to defend against vulnerabilities. Unfortunately, ChatGPT does not fully understand this, as it still uses vulnerability protected mechanisms as an important evaluation indicator in many cases and further reports false positives when detecting vulnerabilities.

\definecolor{darkgreen}{RGB}{100,150,100}
\begin{lstlisting}[caption=Example of false positives in smart contracts due to Protected Mechanism Bias.,label=lst:code1]
contract Personal_Bank {
  (*@\color{darkgreen}// Here is the SafeMath library.@*)
  mapping (address=>uint256) public balances;
  function Deposit(uint256 amount) public payable {
    require(amount > 0, "Amount must be greater than 0");
    (*@\color{darkgreen}// balances[msg.sender] += amount;@*)
    balances[msg.sender] = balances[msg.sender].add(amount);
  }
  function Withdraw(uint256 amount) public payable {
    require(amount > 0, "Amount must be greater than 0");
    balances[msg.sender] = balances[msg.sender].sub(amount);
  }
}
\end{lstlisting}

\begin{center}
\begin{tcolorbox}[colback=gray!10,
                  colframe=black,
                  width=\textwidth,
                  arc=1mm, auto outer arc,
                  boxrule=0.5pt,
                 ]
\noindent
\emph{\textbf{Finding 4:} The detection of certain smart contract vulnerabilities by ChatGPT overemphasizes on protected mechanisms in the code, which may produce false positives.}
\end{tcolorbox}
\end{center}

\definecolor{darkgreen}{RGB}{100,150,100}
\begin{lstlisting}[caption=Example of false positive caused by insufficient understanding of ``msg.value".,label=lst:msgvalue]
function Deposit() public payable{
    balances[msg.sender]+= msg.value;
    Log.AddMessage(msg.sender,msg.value,"Put");
}
\end{lstlisting}

\textbf{Development Intent Bias.} In smart contract, variables such as ``msg.value'' carry specific meanings, signifying the monetary value conveyed by the transaction sender~\cite{solidity_document}. These variables, unlike general variables, are subject to upper limits determined by factors such as the sender's available funds and the total currency supply~\cite{msgvalue}. Consequently, arithmetic operations involving these variables cannot result in overflow concerns. Nevertheless, ChatGPT's comprehension of this distinction is limited. It treats these variables as conventional ones, incorrectly assuming their modifiability, and thus causing false positives regarding integer overflow. As of September 2023, the total amount of ETH flowing in Ethereum is 24.97M~\cite{ether}, equivalent to approximately 405.6 billion US dollars, with an order of magnitude of 10. However, the threshold for \textit{Arithmetic Issues} is 2 to the 256 power~\cite{chen2023healthier}, and its order of magnitude is 77. There exists a significant disparity between these two values. Even if the total ETH supply continues to grow, it is highly improbable that the balance of an account or the amount sent in a transaction approaches the \textit{Arithmetic Issues} threshold under normal circumstances. Listing~\ref{lst:msgvalue} shows the \textit{Deposit} function in a smart contract, and line 2 implements an addition operation. Due to the upper limit of the variable ``msg.value'', this statement does not result in the \textit{Arithmetic Issues} vulnerability. However, GPT-4 believed that the contract has an \textit{Arithmetic Issues} vulnerability.

\definecolor{darkgreen}{RGB}{100,150,100}
\begin{lstlisting}[caption=Example of false positives in smart contracts due to Development Intent Bias.,label=lst:code2]
contract MyContract {
    mapping(address => uint256) public balances;
    function transferAndCheck(address to, uint256 amount) internal returns (bool success) {
        (*@\color{darkgreen}// Check return values.@*)
        require(balances[msg.sender] >= amount, "Insufficient balance");
        (*@\color{darkgreen}// Execute transfer operation.@*)
        balances[msg.sender] -= amount;
        balances[to] += amount;
        emit TransferSuccessful(msg.sender, to, amount);
        success = true;
    }
    event TransferSuccessful(address from, address to, uint256 amount);
    function transferTokens(address to, uint256 amount) public {
        (*@\color{darkgreen}// Calling the transferAndCheck function.@*)
        transferAndCheck(to, amount);
    }
}
\end{lstlisting}

In addition to special variables, some smart contracts may also contain special functions. Listing~\ref{lst:code2} is an example. The \textit{transferAndCheck} function (line 3) is used to check the return value of the transfer and perform the transfer operation, recording relevant events based on the return results. The function \textit{transferTokens} (line 13) only needs to call the \textit{transferAndCheck} function to execute the transfer, without the need to check the return value of the transfer. This approach can make the code more concise and clear, with centralized processing of return value checking. In fact, many token transfer functions (ERC-20~\cite{ERC20}, ERC-721~\cite{ERC721}, etc.) in smart contracts are designed in this way. Therefore, such contracts are usually secure, but GPT may believe that there is a \textit{Unchecked Return Values} vulnerability in function \textit{transferTokens} (line 13) due to its development intent bias.

\begin{center}
\begin{tcolorbox}[colback=gray!10,
                  colframe=black,
                  width=\textwidth,
                  arc=1mm, auto outer arc,
                  boxrule=0.5pt,
                 ]
\noindent
\emph{\textbf{Finding 5:} ChatGPT's analysis of smart contract suffers from development intent bias, resulting in misunderstandings of specific variables and functions, consequently generating false positives for certain vulnerabilities.}
\end{tcolorbox}
\end{center}

\textbf{Interfered by Comments.} To aid comprehension, smart contract code typically includes comments written by developers. The content in these comments is usually an explanation of a certain piece of code. However, ChatGPT, as a dialogue model of natural language processing, cannot distinguish between comments and code itself when analyzing code containing comments. Therefore, it is often interfered with by comments during vulnerability detection. Specifically, when analyzing a piece of code that does not contain vulnerabilities, if the relevant comments contain some logic or statements that contain vulnerabilities, ChatGPT will still determine the code to have vulnerabilities. As shown in Listing~\ref{lst:code1}, a developer originally wrote a ``balances[msg.sender] += msg.value;'' statement in the code with an Integer Overflow vulnerability (line 6), but found the vulnerability after auditing, changed the original statement to a comment, and correctly added the secure ``balances[msg.sender] = balances[msg.sender].add(msg.value);'' statement. However, ChatGPT can misinterpret these fragments, yielding false positives.

\begin{center}
\begin{tcolorbox}[colback=gray!10,
                  colframe=black,
                  width=\textwidth,
                  arc=1mm, auto outer arc,
                  boxrule=0.5pt,
                 ]
\noindent
\emph{\textbf{Finding 6:} ChatGPT's vulnerability detection is influenced by code comments, often interpreting them as actual code, leading to potential false positives.}
\end{tcolorbox}
\end{center}

\textbf{Interference with Dead Code.} Although some code appears to have some vulnerabilities, this does not mean that the execution of this code is unsafe. Because the code fragment containing the vulnerability may be located in a section that will not be executed, such as within an ``if'' conditional statement, but this condition will never be met. In actual execution, such vulnerabilities cannot occur and attackers cannot take the corresponding exploit. However, without the capability to execute code, ChatGPT can misinterpret these fragments, leading to false positives.

\begin{center}
\begin{tcolorbox}[colback=gray!10,
                  colframe=black,
                  width=\textwidth,
                  arc=1mm, auto outer arc,
                  boxrule=0.5pt,
                 ]
\noindent
\emph{\textbf{Finding 7:} Due to the lack of actual execution capability of the code, ChatGPT is prone to false positives due to some dead code.}
\end{tcolorbox}
\end{center}

\noindent\textbf{Distribution.}
In addition, we also counted the proportion of false positives caused by the above reasons in a total of 375 false positives (because each vulnerability generates FPs, some contracts are recalculated) generated by gpt-4, in order to have a more comprehensive understanding of the logic of detecting smart contract vulnerabilities in ChatGPT. The statistical results are shown in Table~\ref{tab:proportion_of_fp}.

\renewcommand{\arraystretch}{1.25}
\begin{table}[ht]
\rowcolors{1}{gray!15}{white}
    \centering
        \caption{The proportion of false positives caused by various reasons.}
        
            \begin{tabular}{l|l|l}
                \hline
                \textbf{Reason for false positive} & \textbf{Number} & \textbf{proportion}\\
                \hline
Protected Mechanism Bias & 124 & 33.1\% \\

Lack of understanding of actual meaning & 191 & 50.9\% \\

Disturbed by comments & 11 & 2.9\% \\
Interference with unexecuted code & 45 & 12.0\% \\
Others & 4 & 1.1\% \\
\hline
        \end{tabular}
        
        \label{tab:proportion_of_fp}
\end{table}
\renewcommand{\arraystretch}{1}

In addition to the above reasons, there are also some false positives caused by other reasons, represented by ``others'' in Table~\ref{tab:proportion_of_fp}. It can be seen that the proportion of false positives is the highest due to a ``lack of understanding of the actual meaning'', indicating that ChatGPT is more likely to generate false positives when detecting smart contract vulnerabilities because it cannot understand the actual meaning of code fragments. On the contrary, among the four reasons we identified, the proportion of false positives caused by ``comment interference'' is the lowest. This is because the smart contracts in the dataset are all from the real world and developers generally do not intentionally leave interfering comments in the code. Nonetheless, this should not undermine the importance of the issue, as it could potentially be exploited by unscrupulous developers. For example, they may strategically insert seemingly secure comments alongside malicious code to evade vulnerability detection of LLMs.

From the perspective of LLMs, ChatGPT produces false positives when detecting smart contract vulnerabilities primarily for the following reasons: 1) Lack of execution context. As an LLM, ChatGPT cannot truly execute code and can only  treat code as text, using semantic analysis to detect vulnerabilities\cite{10.1145/3597503.3639117}. However, for code that only exhibits vulnerabilities under specific execution paths or conditions can easily lead to false positives when analyzed semantically alone. Scenarios such as ``Interference with Dead Code'' and ``Interfered by Comments'' mentioned fall into this category. 2) Insufficient understanding of specialized semantics. Smart contracts employ certain special variables (e.g., msg.value) and functions that differ in meaning and functionality from ordinary variables and functions. Due to the lack of relevant domain knowledge~\cite{NEURIPS2023_1190733f}, LLMs often struggle to accurately comprehend these specialized semantics, leading to erroneous judgments such as ``Development Intent Bias''. 3) Local preference influence. During training, LLMs can form a local preference~\cite{li2024dissecting}. In the field of smart contract vulnerability detection, this can be reflected in focusing too much on certain vulnerability protection mechanisms (e.g., \textit{onlyOwner}), leading to excessive reliance on these mechanisms during detection and resulting in biased evaluations. This reflects the locality preference exhibited by LLMs. 

Overall, the ability of LLMs to perform domain-specific tasks is constrained by the lack of an execution environment, professional knowledge, and contextual understanding capabilities. Future improvements in their performance on such tasks could be improved by integrating domain knowledge and enhancing model architectures.

\textbf{Answer to RQ1:} Overall, the effectiveness of ChatGPT in detecting vulnerabilities in smart contracts does not satisfy practical applications. It has a high recall rate (51.4\% for gpt-3.5-turbo, 83.6\% for gpt-4o and 88.2\% for gpt-4), but low precision (19.7\% for gpt-3.5-turbo, 20.2\% for gpt-4o and 22.6\% for gpt-4), often giving false positives to a smart contract.
	\UseRawInputEncoding
\pdfoutput=1
\section{(RQ2) Comparison with Other Detection Tools}
\label{sec:comparison}

To compare the detection performance of ChatGPT against other tools, we selected 14 smart contract vulnerability detection tools and tested them on the smartbug-cruated dataset. These 14 tools are as follows: Conkas~\cite{conkas}, Oyente~\cite{oyente}, Smartcheck~\cite{smartcheck}, Honeybadger~\cite{honeybadger}, Maian~\cite{maian}, Securify~\cite{securify}, Slither~\cite{slither}, Osiris~\cite{osiris}, Smartian~\cite{smartian}, Confuzzius~\cite{confuzzius}, Sailfish~\cite{sailfish}, Mythril~\cite{consensysmythril}, Solhint~\cite{solhint}, AChecker~\cite{ghaleb2023achecker}. There are three reasons for choosing these 14 tools: 1) Their source code are available. 2) They support to detect at least one type of vulnerability within DASP10. 3) They are published on top venues or wildly used by smart contract community.

\subsection{Effectiveness}
Table~\ref{tab:comparison} shows the detection performance of different tools in the smartbug-cruated dataset. Each tool's value is represented by F-measure for a specific vulnerability type. Vulnerabilities beyond a tool's detection scope are marked with ``/''. Additionally, the two vulnerabilities \textit{Bad Randomness} and \textit{Short Address Attack} are not supported by all the 14 tools, so they are omitted in Table~\ref{tab:comparison}.

\begin{table}[ht]
    \centering
    \small
        \caption{Comparison of detection performance between ChatGPT and other tools.}
        
            \begin{tabular}{l|p{1.6cm}|p{1.1cm}|p{1.5cm}|p{1.5cm}|p{1cm}|p{1cm}|p{1.2cm}}
            \hline
            \textbf{ } &
            \textbf{Reentrancy} & \textbf{Access Control} & \textbf{Arithmetic Issues} & \textbf{Unchecked Return Values} & \textbf{Denial of Service} & \textbf{Front Running} & \textbf{Time Manipulation}\\
            \hline
            \textbf{Conkas} & 44.0\% & / & 16.8\% & 91.1\% & / & 7.7\% & 27.0\% \\
            \textbf{Oyente} & \textbf{87.5\%} & 0.0\% & 23.0\% & / & 3.7\% & / & 0.0\% \\
            \textbf{Smartcheck} & 84.1\% & 18.2\% & 6.7\% & \textbf{92.9\%} & 7.7\% & / & 33.3\% \\
            \textbf{Honeybadger} & 76.0\% & / & 0.0\% & / & / & / & / \\
            \textbf{Maian} & / & \textbf{51.9\%} & / & / & / & / & / \\
            \textbf{Securify} & 72.0\% & 8.7\% & / & 91.7\% & / & 7.1\% & / \\
            \textbf{Slither} & 87.0\% & 23.1\% & / & 72.7\% & 10.0\% & / & \textbf{60.0\%} \\
            \textbf{Osiris} & 64.4\% & / & 28.6\% & / & 6.7\% & / & 36.4\% \\
            \textbf{Smartian} & 30.8\% & 18.2\% & 60.0\% & 84.9\% & 0.0\% & / & 23.5\% \\
            \textbf{Confuzzius} & 76.4\% & 46.7\% & \textbf{61.1\%} & 86.3\% & / & 10.5\% & / \\
            \textbf{Sailfish} & 71.7\% & / & / & / & 5.9\% & 5.9\% & / \\
            \textbf{Mythril} & 37.2\% & 13.3\% & 51.4\% & 67.5\% & / & 0.0\% & / \\
            \textbf{Solhint} & 51.3\% & 27.5\% & / & 72.7\% & 0.0\% & / & 22.7\% \\
            \textbf{AChecker} & / & 50.0\% & / & / & / & / & / \\
            \hline
            \hline
            \textit{Ave.} & 65.2\% & 25.8\% & 30.6\% & 82.5\% & 4.9\% & 4.7\% & 29.0\% \\
            \hline
            \hline
            \textbf{ChatGPT} & 49.8\% & 28.4\% & 35.7\% & 67.9\% & \textbf{12.8\%} & \textbf{15.9\%} & 28.4\% \\
            \hline
        \end{tabular}
        
        \label{tab:comparison}
\end{table}

The results in Table~\ref{tab:comparison} reveal that ChatGPT's detection capabilities, particularly for vulnerabilities, such as  \textit{Reentrancy} and \textit{Access Control} is not as good as other current tools. However, the performance of ChatGPT is relatively optimal on \textit{Denial of Service} and \textit{Front Running}. This superiority likely derives from the inherent complexity that characterizes these vulnerabilities. \textit{DoS} attacks usually leads to contract failure because of excessive resource use under specific conditions. For example, an attacker might flood the contract with requests, causing it to run out of gas and the failure of execution. \textit{Front Running} involves gaining access to transaction details before they are officially submitted and then taking advantage of this information to maximize profits by miners. Due to intricate nature of the two vulnerabilities, it is challenging to define rules to cover all the patterns for them based on traditional program analysis methods. However, ChatGPT's ability to understand code semantics enhances its performance in vulnerability detection. Unlike many program analysis tools that rely on pattern matching, ChatGPT can grasp complex logical structures, making it more effective at detecting vulnerabilities such as \textit{DoS} and \textit{Front Running}. Even so, its F-measure is not very high.
It's notable that while 14 selected tools perform better in some types of vulnerability, they are completely unable to detect vulnerability outside their predefined scopes. On the contrary, ChatGPT is not limited to specific vulnerability types and can detect a wider range of vulnerabilities, although success rates may vary. This indicates that ChatGPT has a good breadth in detecting smart contract vulnerabilities, but its accuracy needs to be improved.

\begin{center}
\begin{tcolorbox}[colback=gray!10,
                  colframe=black,
                  width=\textwidth,
                  arc=1mm, auto outer arc,
                  boxrule=0.5pt,
                 ]
\noindent
\emph{\textbf{Finding 8:} Among the nine vulnerabilities that we have examined, ChatGPT demonstrates a notable capability by successfully detecting all nine types of vulnerabilities, showcasing its comprehensive coverage in vulnerability detection.}
\end{tcolorbox}
\end{center}

\begin{center}
\begin{tcolorbox}[colback=gray!10,
                  colframe=black,
                  width=\textwidth,
                  arc=1mm, auto outer arc,
                  boxrule=0.5pt,
                 ]
\noindent
\emph{\textbf{Finding 9:} In comparison to the other 14 tools, ChatGPT demonstrates better detection performance in two vulnerabilities: \textit{Denial of Service} and \textit{Front Running}. However, in 3 out of 7 vulnerabilities, the detection performance of ChatGPT is inferior to the average level of other tools.}
\end{tcolorbox}
\end{center}

\subsection{Efficiency}
\label{subsec:efficiency}
The efficiency of these tools is a crucial factor in assessing their practical utility in the field of smart contract security. Therefore, we recorded the time it took for each tool and GPT models to conduct experiments on the entire dataset. It is used as the metric to evaluate their efficiency. Table~\ref{tab:time} shows the time performance of various tools and ChatGPT for vulnerability detection on the dataset. In general, the efficiency of ChatGPT is good. gpt-3.5-turbo, gpt-4o and gpt-4 completed the detection of the whole dataset in 9 minutes and 21 seconds, 9 minutes and 55 seconds, 11 minutes and 9 seconds, respectively. It's worth highlighting that the efficiency demonstrated by all three gpt models surpassed that of the other 11 specialized tools (78.6\%) listed in Table~\ref{tab:time}, only inferior to that of \textit{SmartCheck}, \textit{Solhint} and \textit{AChecker}.

\begin{table}[ht]
    \centering
    \caption{Comparison of time performance between ChatGPT and other tools in detecting 142 smart contracts on SmartBugs Dataset.}
    
        \begin{tabular}{p{2.5cm}|l|l}
            \hline
            \textbf{Tools} & \textbf{Time(h:m:s)} & \textbf{errors}\\
            \hline
            Conkas & 1:05:23 & 0 \\
            Oyente & 0:12:44 & 0 \\
            Smartcheck & 0:05:31 & 0 \\
            Honeybadger & 1:09:38 & 0 \\
            Maian & 2:53:22 & 0 \\
            Securify & 0:52:04 & 0 \\
            Slither & 0:38:50 & 0 \\
            Osiris & 1:11:14 & 0 \\
            Smartian & 0:22:44 & 0\\
            Confuzzius & 0:11:23 & 7 \\
            Sailfish & 0:25:12 & 27\\
            Mythril & 7:10:02 & 4 \\
            Solhint & 0:04:43 & 0 \\
            AChecker & 0:06:49 & 2 \\
            \hline
            gpt-3.5-turbo & 0:09:21 & 5 \\
            gpt-4o & 0:09:55 & 5 \\
            gpt-4 & 0:11:09 & 2 \\
            \hline
        \end{tabular}
    
    \label{tab:time}
\end{table}

\begin{center}
\begin{tcolorbox}[colback=gray!10,
                  colframe=black,
                  width=\textwidth,
                  arc=1mm, auto outer arc,
                  boxrule=0.5pt,
                 ]
\noindent
\emph{\textbf{Finding 10:} In the realm of vulnerability detection, gpt-3.5-turbo, gpt-4o and gpt-4 all excel in terms of efficiency, outperforming 11 out of 14 other tools.}
\end{tcolorbox}
\end{center}

\textbf{Answer to RQ2:} Compared to other tools, ChatGPT can detect more diverse types of vulnerability, but its detection performance score is not very ideal. In addition, ChatGPT outperforms most tools in terms of efficiency.
	\UseRawInputEncoding
\pdfoutput=1
\section{(RQ3) Limitations of ChatGPT in Detecting Vulnerabilities}
\label{sec:limitations}
Although ChatGPT can be used to detect vulnerabilities in smart contracts, it also exhibits many limitations. The first limitation is the uncertainty output. For example, under the same input conditions, ChatGPT may produce different detection results, thus introducing a certain degree of variability. This uncertainty can be derived from the design and training methods of the ChatGPT model.

In addition, the length of the smart contract will also have an impact on the detection performance of ChatGPT. When the contract is too long, ChatGPT may encounter detection failures. This is because ChatGPT is a pretrained language model, and its input limitations are usually limited by the maximum input length of the model. As shown in Table~\ref{tab:time}, 5 contracts failed to be detected due to the limitation of the input token of the GPT model, while 10 other tools (71.4\%) did not have any failure cases on the dataset. When the contract exceeds the limit on the length of the code, ChatGPT may not be able to fully understand and process the context of the entire contract, resulting in a detection failure.

\subsection{Uncertainty}
As a language model-based AI model, ChatGPT determines the vulnerabilities in input smart contracts probabilistically. Therefore, even with the same smart contract inputs, ChatGPT might generate divergent vulnerability detection results in different runs. Such variability can be attributed to the model's internal state, randomness, and minuscule input variations.

\begin{figure}[ht]
    \centering
    \includegraphics[width=280pt]{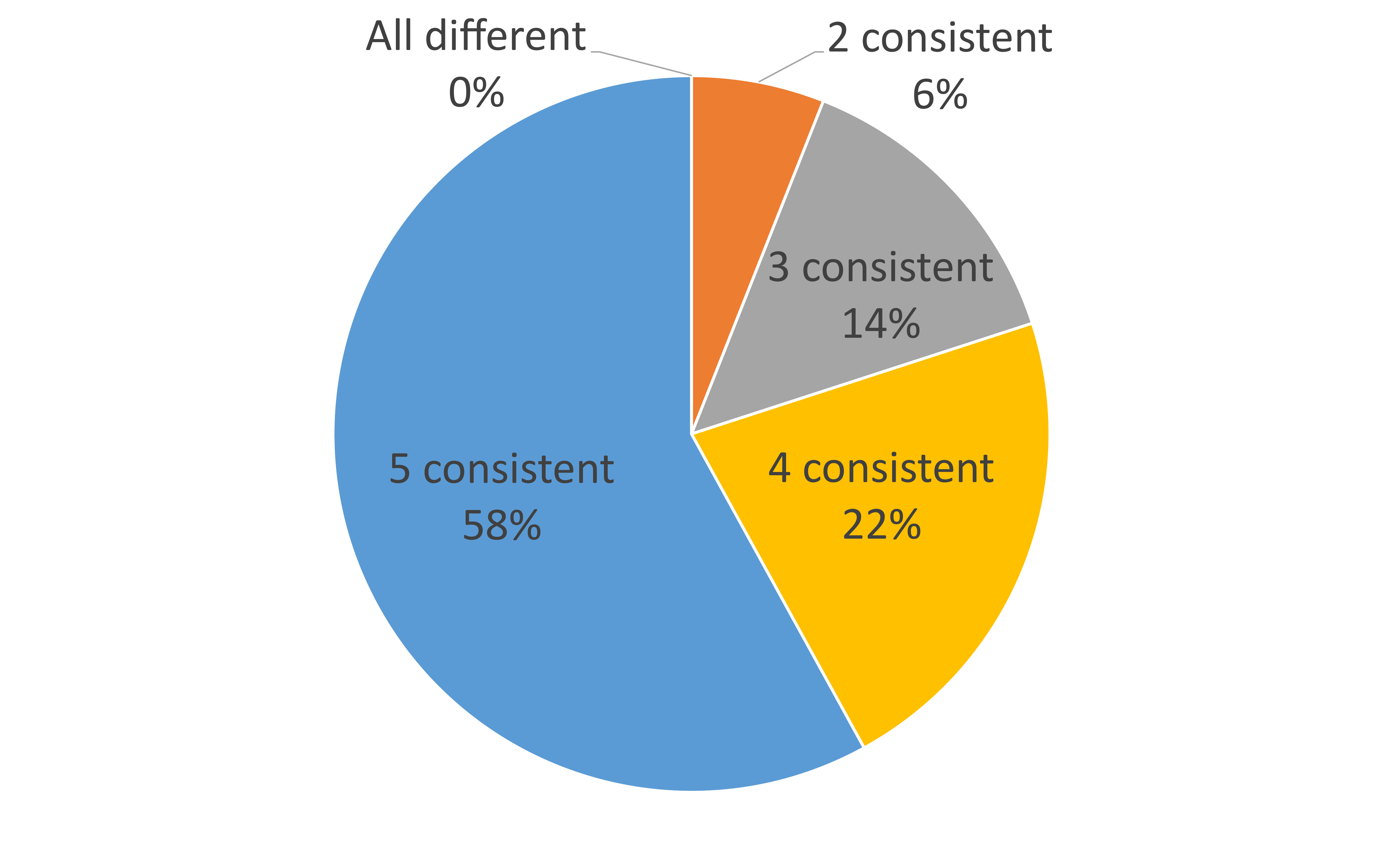}
    \caption{The uncertainty test results of ChatGPT. Each part displays the number of contracts with consistent results, for example, 5 consistent represents the number of contracts with consistent results from all 5 tests.}
    \label{fig:fig5}
\end{figure}

To delve deeper into ChatGPT's result uncertainty when detecting smart contract vulnerabilities, we randomly selected 50 smart contracts from the dataset and conducted five rounds of vulnerability detections on each of the 50 smart contracts using the gpt-4 model. We then examined the consistency across the five outcomes for every contract. To some extent, for a smart contract, the more consistent the number of results, the stronger the certainty of ChatGPT on the result, and vice versa. The uncertainty of vulnerability detection results of ChatGPT on smart contracts is shown in Figure~\ref{fig:fig5}. Each block of Figure~\ref{fig:fig5} displays the number of contracts with consistent results (e.g., 5 consistent represents the number of contracts with consistent results from all 5 tests). In this experiment, we set the confidence level of the results generated by the GPT model to 80\% (i.e., if no less than 4 out of 5 results are consistent, we believe that this result is reliable). Among the 50 randomly selected smart contracts, ChatGPT only gave reliable results on 80\% of the contracts, while the remaining 20\% were unreliable. It can be seen that, in some cases, regardless of whether the conclusion is right or wrong, ChatGPT cannot guarantee the certainty of the conclusion it provides.

GPTScan~\cite{sun2023gpt} employs two strategies to mitigate the ambiguity in the responses generated by ChatGPT: adjust the model temperature and employ the ``mimic-in-the-background'' prompting technique. The former involves setting the temperature parameter of the GPT model to 0, thereby ensuring consistent outputs for identical prompts. The latter employs a prompt methodology, leveraging the GPT system's prompt instruction model to simulate responding to a query five times in the background, ultimately delivering the most frequently recurring response to enhance consistency. Both of these methods contribute to the refinement of ChatGPT's output results. Hence, we conducted comparative experiments using the same settings and datasets to assess their respective impacts, with detailed results presented in Table~\ref{tab:uncertainty_comp}.

It can be seen that for gpt-3.5-turbo, while both methods can produce some improvements in the detection performance of specific vulnerabilities, the overall efficacy has diminished. In contrast, in the case of gpt-4, these two methods can marginally enhance vulnerability detection performance while maintaining the certainty of the results. This discrepancy may be attributed to the inherent improvements in the gpt-4 architecture, which better leverages these strategies to balance performance and consistency, unlike the earlier gpt-3.5-turbo model.

\begin{table}[htbp]
\centering
\caption{Results after using methods to reduce uncertainty, where ``temp-adj'' represents adjusting the temperature to 0, ``MITB'' represents mimic in-the-background prompting.}
\resizebox{0.8\textwidth}{!}{%
\begin{tabular}{@{}lcccccc@{}}

\toprule
 & \multicolumn{3}{c}{gpt-3.5-turbo} & \multicolumn{3}{c}{gpt-4} \\ \cmidrule(l){2-7} 
 & original & temp-adj & MITB & original & temp-adj & MITB \\ \midrule
Reentrancy & 33.8 & $54.5 \uparrow$ & $45.7 \uparrow$ & 49.8 & $59.8 \uparrow$ & $54.1 \uparrow$ \\
Access Control & 23.9 & $29.5 \uparrow$ & $28.6 \uparrow$ & 28.4 & $32.2 \uparrow$ & $31.7 \uparrow$ \\
Arithmetic Issues & 34.8 & 21.1 & 30.8 & 35.7 & $44 \uparrow$ & $40 \uparrow$ \\
Unchecked Return Values & 59 & 41.6 & 34.7 & 67.9 & $68.2 \uparrow$ & 64.4 \\
Denial of Service & 20.8 & 5.9 & 16.7 & 12.8 & $19.4 \uparrow$ & $16.9 \uparrow$ \\
Bad Randomness & 35 & 7.4 & 27.3 & 57 & $60.9 \uparrow$ & 52.2 \\
Front Running & 7.7 & 0 & 0 & 15.9 & 7.6 & $30.8 \uparrow$ \\
Time manipulation & 19 & $28.6 \uparrow$ & $28.6 \uparrow$ & 28.4 & 24 & $29.6 \uparrow$ \\
Short Address Attack & 0 & 0 & 0 & 0 & 0 & $22.2 \uparrow$ \\ \midrule
Avg. & 26.0 & 21.0 & 23.6 & 32.9 & $35.1 \uparrow$ & $38.0 \uparrow$ \\ \bottomrule
\end{tabular}%
}
\label{tab:uncertainty_comp}
\end{table}

\subsection{Context Length}
ChatGPT has token restrictions for contextual input, and it may face detection failure when detecting smart contracts due to the excessively long code length. Despite the token limit guidance provided by GPT, real-world scenarios often introduce various complexities, such as comments, statement lengths, and other factors that can impact the effective code length that the model can handle. Therefore, an experiment was conducted by us to approximate the maximum code size they can accommodate. We calculated the token length of the prompt designed in this study and analyzed the smart contract code that failed detection to explore the threshold code length that the model can detect.

For gpt-4, the maximum number of tokens it can accept is 8192. In this study, the number of tokens occupied by our designed prompt is 107, which means that the maximum length occupied by the input code can be 8085 tokens. We have reviewed two smart contracts that failed detection, with lengths of 2470 and 897 lines of code, and file sizes of 94.4KB and 35.5KB, respectively. Among all successfully detected smart contracts, the longest code length is 771 lines, corresponding to a file size of 27KB. We are attempting to extend the 771 line code by adding some useless comments or code fragments. When the file size reaches 33KB and the number of code lines reaches 857, gpt-4 detection cannot proceed properly. Similarly, we conducted the same experiment on gpt-3.5 and found that the maximum smart contract size that the gpt-3.5 model can detect is 16KB. Its value is much lower than gpt-4 because the maximum number of tokens that the gpt-3.5 model can accept is twice that of gpt-4. Therefore, we can draw a rough conclusion that under the experimental conditions of this study, the maximum code size that the gpt-4 model and gpt-3.5 model can detect is around 33KB in file size and 16KB in file size, respectively.

\textbf{Answer to RQ3:} ChatGPT exhibits certain limitations in detecting vulnerabilities in smart contracts, notably in the uncertainty of detection conclusions and the inability to detect lengthy contracts due to token limitations in the model.
	\UseRawInputEncoding
\pdfoutput=1
\section{Discussion}
\label{sec:discussion}
In the course of using ChatGPT for the detection of vulnerability in smart contracts, our paper uncovered intriguing phenomena, led to insightful reflections and relevant experiments. First, we delved into ChatGPT's capability to grasp vulnerabilities and showcased the potential for enhancing its detection prowess by injecting external knowledge through illustrative examples. Following that, we embarked on an exploration of multi-round conversation experiments to ascertain whether ChatGPT's multi-round interactions with users wield influence over its vulnerability detection performance. Additionally, we orchestrated code poisoning attack experiments to assess ChatGPT's response to protected mechanisms within the code. Lastly, we orchestrated character obfuscation attack experiments to investigate whether ChatGPT relies solely on character-level semantics for code analysis.

\subsection{Vulnerability learning}
Based on the experimental results in Section~\ref{sec:effectiveness}, we found that the effectiveness of ChatGPT in detecting smart contract vulnerabilities is limited. As a pre-trained large language model, ChatGPT may only have limited knowledge of smart contract vulnerabilities. In other words, some new vulnerabilities that appear after a specific date (the date for the corpus used to pre-train the GPT model) cannot be detected by ChatGPT. In addition, if there is less information about a certain vulnerability in the pre-trained data, it will also lead to poor detection performance of ChatGPT for the vulnerability. Therefore, we attempt to identify vulnerabilities with poor ChatGPT detection performance and improve the detection performance of such vulnerabilities by injecting external knowledge. For example, \textit{Price Oracle Manipulation} is a kind of vulnerability with a poor detection effect of ChatGPT. The price oracle machine is a component that can provide asset price data on the chain. Many DeFi projects will use price oracle machine contracts to obtain prices. The \textit{Price Oracle Manipulation} is to manipulate the price oracle machine through various means to return wrong price data, thereby affecting DeFi projects that rely on these data and causing financial losses.

\begin{figure}[ht]
    \centering
    \includegraphics[width=280pt]{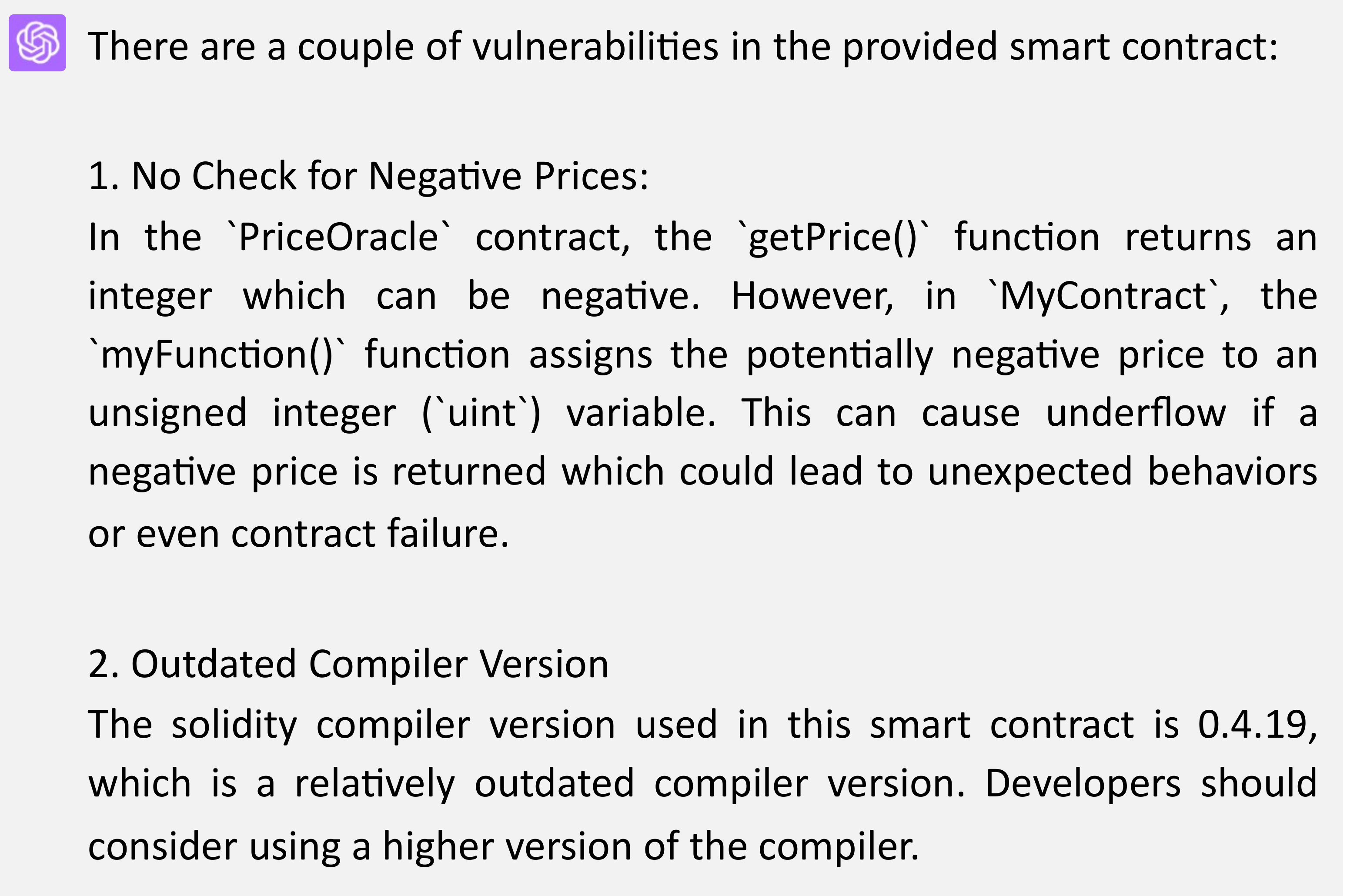}
    \caption{GPT-4's original detection results of smart contracts containing \textit{Price Oracle Manipulation} vulnerabilities.}
    \label{fig:fig8}
\end{figure}

Listing~\ref{lst:code3} is a smart contract containing a \textit{Price Oracle Manipulation} vulnerability. The \textit{MyContract} contract obtains (line 9) on chain price data through the PriceOracle contract. The priceFeed address can be set in PriceOracle (line 5). Attackers can set a malicious priceFeed address through governance proposals or other methods. The malicious Oracle machine contract can manipulate the price at will, so that key operations in MyContract will produce unexpected results. We successively gave five different smart contracts containing this vulnerability to ChatGPT and asked them to detect it. ChatGPT did not detect this vulnerability once. Figure ~\ref{fig:fig8} shows the detection results of a certain contract for this vulnerability. This detection result points out two secondary problems in the smart contract, while ignoring the main \textit{Price Oracle Manipulation} vulnerability.

\definecolor{darkgreen}{RGB}{100,150,100}
\begin{lstlisting}[caption=Example of smart contract containing \textit{Price Oracle Manipulation} vulnerability.,label=lst:code3]
contract PriceOracle {
  AggregatorV3Interface public priceFeed;
  constructor(address _priceFeed) {
    priceFeed = AggregatorV3Interface(_priceFeed); }
  function getPrice() public view returns (int) {
    return priceFeed.latestRoundData().answer;}
}
contract MyContract {  
  PriceOracle priceOracle;
  function myFunction() external {
    uint price = priceOracle.getPrice(); 
    (*@\color{darkgreen}// Use the variable to perform some key operations.@*)}
}
\end{lstlisting}

The efficacy of employing pre-existing knowledge for directly detecting \textit{Price Oracle Manipulation} vulnerability has been unsatisfactory. Therefore, it is worth exploring the integration of external knowledge as contextual hints within prompts to enhance performance. A key factor is finding the appropriate amount of hints to guide ChatGPT effectively. Therefore, we attempt to improve the detection performance of this vulnerability by injecting external information, and our approach involved categorizing the information in the prompts into three levels to assess ChatGPT's performance (GPT-4 was selected for this experiment): 1) No Information Provided: In the initial tests without specific prompts, ChatGPT failed to identify \textit{Price Oracle Manipulation} vulnerabilities present in 5 smart contracts. 2) Vulnerability Name Only: When the term ``Price Oracle Manipulation Vulnerability'' was explicitly mentioned in the prompt, ChatGPT successfully detected the vulnerability in 2 out of 5 contracts. 3) Vulnerability Name and Code Example: By adopting a technique of providing a few-shot example prompt, which incorporates both the vulnerability name and a code sample, ChatGPT was able to accurately identify the \textit{Price Oracle Manipulation} vulnerability across all 5 smart contracts. Figure 5 shows one of the ChatGPT detection results for this vulnerability after adding external knowledge. This finding suggests that ChatGPT has the ability to incorporate and learn from external knowledge effectively, making it applicable for subsequent detection tasks.


\begin{figure}[ht]
    \centering
    \includegraphics[width=280pt]{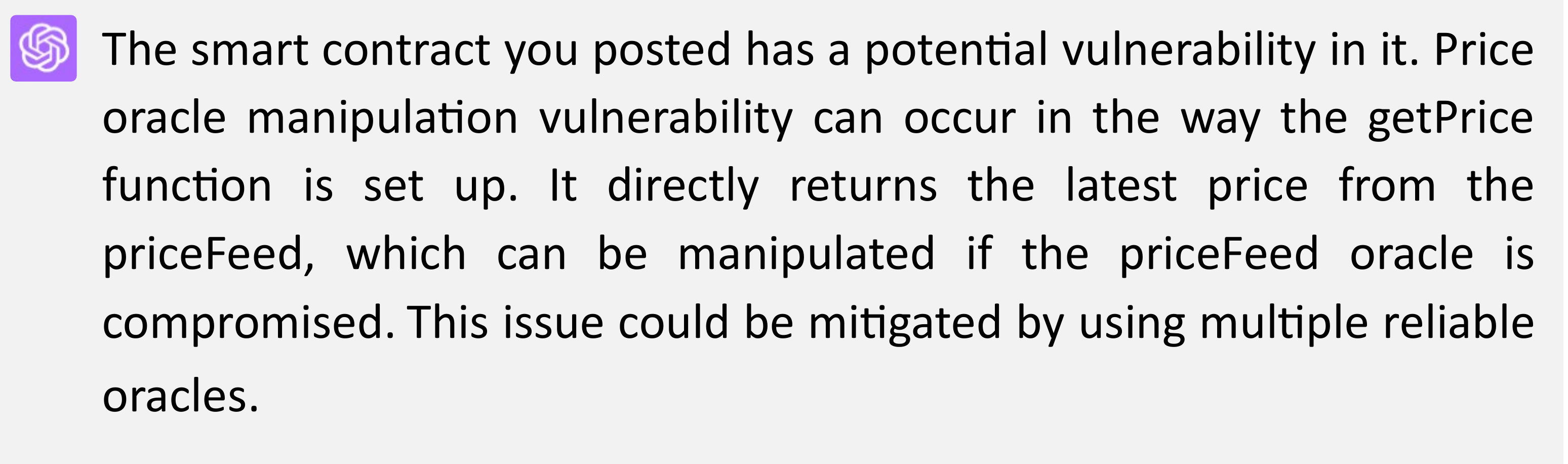}
    \caption{GPT-4's improved detection results of smart contracts containing \textit{Price Oracle Manipulation} vulnerabilities.}
    \label{fig:fig9}
\end{figure}
\
\subsection{Multi-round Conversation}
\begin{figure}[ht]
    \centering
    \includegraphics[width=280pt]{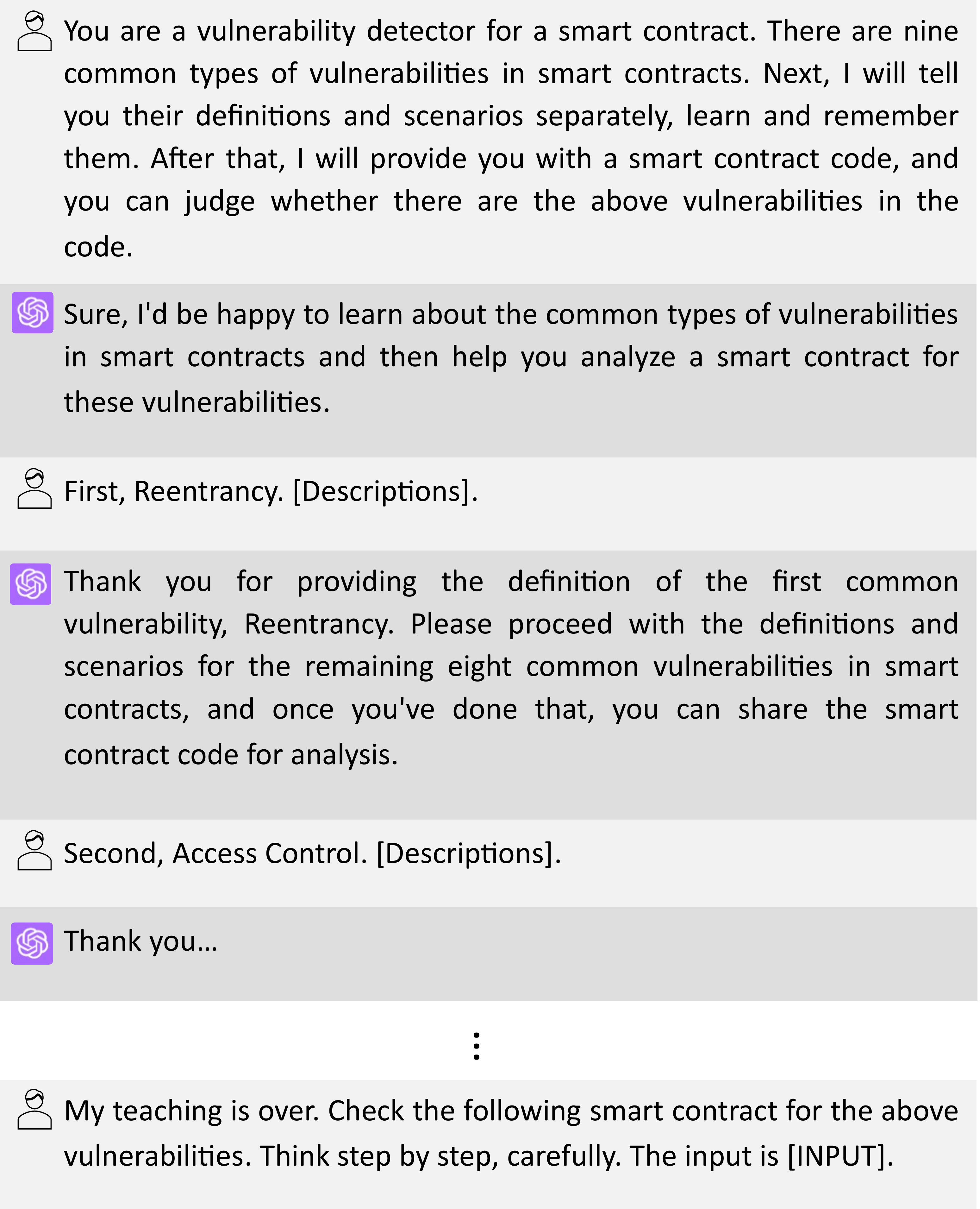}
    \caption{Schematic diagram for multiple rounds of conversation with ChatGPT.}
    \label{fig:fig13}
\end{figure}

\begin{figure}[ht]
    \centering
    \includegraphics[width=0.9\columnwidth]{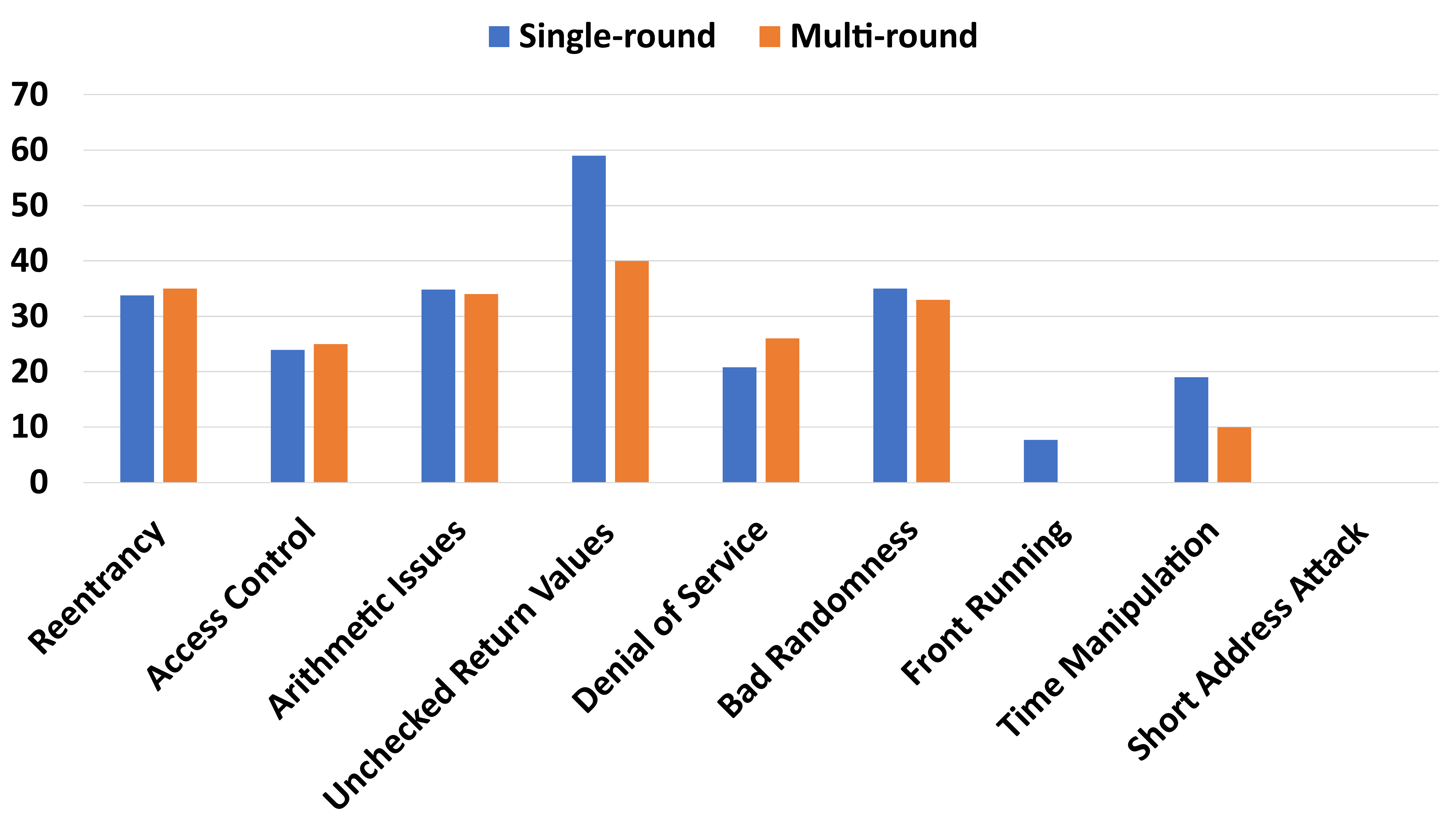}
    \caption{Experimental results of multi-round conversation.}
    \label{fig:fig11}
\end{figure}

Due to ChatGPT's ability of multiple rounds conversations, we attempted to redo the experiment in this way and add more information during each round of conversations. Figure~\ref{fig:fig13} is an example of a multi-round conversation. We conducted nine successive rounds of conversations with gpt-3.5-turbo. In every interaction, in addition to the name and alias of each vulnerability, we also provide definitions of the vulnerability (information is sourced from the official DASP10 website~\cite{dasp2018}). ChatGPT acknowledged its awareness of this information upon receiving it in each round of the conversation. Subsequently, we presented a contract source code to ChatGPT for assessment after the conversation rounds. A comparative analysis between the results of this new method and the previous experiment is shown in Figure ~\ref{fig:fig11}.

The experimental results reveal that there is no obvious enhancement in the detection performance of ChatGPT after multi-round conversation. This observation suggests that the mode of interaction has a limited influence on ChatGPT's capacity to detect vulnerabilities. Compared to the single-round conversation-based experiments, the multi-round conversation-based approach introduced specific vulnerability definitions, thereby furnishing additional information to ChatGPT. Despite this augmentation, upon inspecting the graphical data, it becomes evident that the detection performance scores for certain vulnerability groups, such as ``Arithmetic Issues'' and ``Unchecked Return Values'', exhibit a decline rather than an increase. From the perspective of an LLM, several reasons may contribute to this phenomenon. Firstly, while multiple rounds of conversation inject additional external information into the prompt, an overload of information may confuse the model, making it challenging to effectively process a large amount of detailed data. Secondly, external information may be inconsistent or difficult to integrate with ChatGPT's existing knowledge base, resulting in knowledge conflicts. Finally, over-specialization in specific information may diminish the model's ability to generalize across different contexts, thereby impacting vulnerability detection performance. Hence, when using LLMs for tasks, it is crucial to carefully consider the manner and quantity of information provided to mitigate adverse effects on model performance.

\subsection{Code Poisoning}
As elucidated in Section~\ref{subsec:fp_analysis}, ChatGPT has the risk of generating false positives for vulnerabilities in smart contracts due to the absence of protected mechanisms within the code. From the perspective of attackers, they may explore methods to circumvent ChatGPT's vulnerability detection. For example, they could deceive ChatGPT by incorporating protected mechanisms (e.g., \textit{SafeMath}), leading the models to overlook vulnerabilities during detection. In addition, attackers have the ability to incorporate permission control mechanisms, such as the \textit{onlyOwner} modifier, into contracts harboring \textit{Access Control} vulnerability, thus circumventing detection by ChatGPT. To substantiate this notion from an attack-oriented point of view, we conducted an evaluation. Specifically, in all contracts containing \textit{Arithmetical Issues} vulnerability, we artificially added the \textit{SafeMath} library without calling it (i.e., we only added ``import SafeMath'' to the contract without any other modification, which would not change the logic of the contract). In parallel, we have integrated the \textit{onlyOwner} modifier into all contracts identified with \textit{Access Control} vulnerability where said modifiers were not used. Under this condition, we let the GPT model detect vulnerabilities in these contracts again. In this case, ChatGPT is likely to falsely report the smart contract as negative due to protected mechanism bias.

\begin{table}[htbp]
    \centering
    \subfloat[After adding the \textit{SafeMath} library, the detection results of contracts containing \textit{Arithmetic Issues}.]{\label{tab:AI_poison}
        \begin{tabular}{l|l}
            \hline
            \textbf{Contract Name} & \textbf{Result} \\
            \hline
            BECToken.sol & $\times$ \\
            \hline
            insecure\_transfer.sol & $\checkmark$ \\
            \hline
            integer\_overflow\_1.sol & $\checkmark$ \\
            \hline
            integer\_overflow\_add.sol & $\checkmark$ \\
            \hline
            integer\_overflow\_benign\_1.sol & $\times$ \\
            \hline
            mapping\_sym\_1.sol & $\checkmark$ \\
            \hline
            integer\_overflow\_minimal.sol & $\times$ \\
            \hline
            integer\_overflow\_mul.sol & $\checkmark$ \\
            \hline
            multitx\_multifunc\_feasible.sol & $\times$ \\
            \hline
            multitx\_onefunc\_feasible.sol & $\checkmark$ \\
            \hline
            overflow\_simple\_add.sol & $\checkmark$ \\
            \hline
            overflow\_single\_tx.sol & $\checkmark$ \\
            \hline
            timelock.sol & $\times$ \\
            \hline
            token.sol & $\times$ \\
            \hline
            tokensalechallenge.sol & $\circ$ \\
            \hline
            \end{tabular}
    }    \hfill
    \subfloat[After adding the \textit{onlyOwner} modifier, the detection results of contracts containing \textit{Access Control}.]{\label{tab:AC_poison}
        \begin{tabular}{l|l}
            \hline
            \textbf{Contract Name} & \textbf{Result} \\
            \hline
            arbitrary\_location\_write\_simple.sol & $\times$ \\
            \hline
            FibonacciBalance.sol & $\circ$ \\
            \hline
            incorrect\_constructor\_name1.sol & $\times$ \\
            \hline
            incorrect\_constructor\_name2.sol & $\checkmark$ \\
            \hline
            incorrect\_constructor\_name3.sol & $\times$ \\
            \hline
            mapping\_write.sol & $\checkmark$ \\
            \hline
            multiowned\_vulnerable.sol & $\times$ \\
            \hline
            mycontract.sol & $\times$ \\
            \hline
            parity\_wallet\_bug\_1.sol & $\circ$ \\
            \hline
            parity\_wallet\_bug\_2.sol & $\circ$ \\
            \hline
            phishable.sol & $\checkmark$ \\
            \hline
            proxy.sol & $\times$ \\
            \hline
            rubixi.sol & $\times$ \\
            \hline
            simple\_suicide.sol & $\times$ \\
            \hline
            unprotected0.sol & $\times$ \\
            \hline
            wallet\_02\_refund\_nosub.sol & $\checkmark$ \\
            \hline
            wallet\_03\_wrong\_constructor.sol & $\times$ \\
            \hline
            wallet\_04\_confused\_sign.sol & $\times$ \\
            \hline
            \end{tabular}
    }
    \caption{Experimental results of code poisoning.}
    \label{tab:code_poison}
\end{table}

        

Table~\ref{tab:code_poison} illustrates the results of the code poisoning experiment. The symbol ``$\times$'' denotes contracts that failed the code poisoning test, which implies that while ChatGPT could have originally identified vulnerabilities within these contracts, it was unable to do so after the protected mechanism was incorporated. On the other hand, contracts that successfully underwent the code poisoning experiment are indicated by the ``$\checkmark$'' symbol. Furthermore, contracts that exhibited an inability to detect vulnerabilities both before and after code poisoning are represented by the ``$\circ$'' symbol. In particular, these results reveal that approximately 46.7\% of the smart contracts were influenced by the introduction of \textit{SafeMath} library and 61.1\% by that of \textit{onlyOwner} modifier. It should be noted that this experiment, while a simple exploration of code poisoning, underscored the potential for more sophisticated mechanisms to bypass GPT-based detection. It is possible for this to be exploited for malicious purposes.

\subsection{Code Confusion Attack}
Since code comments can potentially disrupt the functioning of large models (as discussed in Section ~\ref{subsec:fp_analysis}), we hypothesize that GPT analyzes code based on its character-level semantics. To explore this hypothesis, we conducted character-obfuscation attack experiments. In this series of experiments, we introduced deliberate obfuscation into each contract within our dataset. Specifically, we applied obfuscation rules to all identifiers within the contracts, including function names and variable names. These rules involved replacing certain characters with visually similar glyphs (i.e., ``l'' with ``1'', ``o'' with ``0'', ``i'' with ``l'', ``s'' with ``5'', ``g'' with ``9''), and if the replaced identifier began with a number, we prepended an underscore before the number. The experimental results following this obfuscation process are presented in Table ~\ref{tab:obfuscation_comparison}.

\begin{table}[ht]
    \centering
    \caption{Comparison of Vulnerability detection results (F1 score) before and after char obfuscation.}
        \begin{tabular}{l|c|c}
            \hline
            \textbf{Vulnerability} & \textbf{Raw} & \textbf{Obfuscation} \\
            \hline
            Reentrancy & 49.8\% & 64.2\% \\
            Access Control & 28.4\% & 23.5\% \\
            Arithmetic Issues & 35.7\% & 51.6\% \\
            Unchecked Return Values & 67.9\% & 69.1\% \\
            Denial of Service & 12.8\% & 17.9\% \\
            Bad Randomness & 57.0\% & 35.0\% \\
            Front Running & 15.9\% & 8.0\% \\
            Time Manipulation & 28.4\% & 20.5\% \\
            Short Address Attack & 8.4\% & 0.0\% \\
            \hline
            Average & 33.8\% & 32.2\% \\
            \hline
        \end{tabular}
    
    \label{tab:obfuscation_comparison}
\end{table}

After applying character obfuscation, the detection performance of only four vulnerabilities (44.4\%) decreased, indicating susceptibility to obfuscation attacks. Surprisingly, the detection performance of the remaining five vulnerabilities remained relatively stable or even showed slight improvements. Although our study shows that character obfuscation has varying effects on different vulnerabilities and a limited impact on the overall average performance of vulnerability detection, it is important to note that the underlying mechanisms explaining these observations are beyond the scope of this paper. Further in-depth research is needed to explore these underlying principles.
	\UseRawInputEncoding
\pdfoutput=1
\section{Related Work}
\label{sec:related}
\subsection{Vulnerability Detection}
In recent years, there have been many related works on vulnerability detection in smart contracts. A variety of tools and approaches have been used to pinpoint vulnerabilities in the ecosystem of smart contracts. A prevalent methodology is static verification, which evaluates the source code or bytecode without running it~\cite{brent2020ethainter,brent2018vandal,feist2019slither,kalra2018zeus,tsankov2018securify}. On the other hand, dynamic analysis provides more profound discernment by surveying smart contracts during execution. Automated tools, such as fuzzing~\cite{grieco2020echidna,jiang2018contractfuzzer,wustholz2020harvey} can generate input to comprehensively assess the performance of the system. Symbolic execution~\cite{consensysmythril,frank2020ethbmc,luu2016making,mossberg2019manticore} and formal verification~\cite{permenev2020verx,so2019verismart} are commended for their effectiveness, although formal verification characteristically requires specifications given by the user. Furthermore, Zheng et al.~\cite{zheng2023reentrancy} investigated the effectiveness of various current tools in detecting reentry vulnerabilities and summarized the reasons for their false positives. In summary, automated tools and techniques can be used to uncover potential vulnerabilities in smart contracts. Recent studies have pioneered new techniques for identifying profitable manipulations and security threats in decentralized finance (DeFi) systems. Novel methodologies utilizing dynamic program analysis from Qin et al.~\cite{qin2023blockchain,qin2022quantifying} can automatically replicate and synthesize money-making transactions on blockchains in real time. Such techniques could be deployed after smart contracts are released to serve as on-chain intrusion monitoring. Other works like DeFiPoser~\cite{zhou2021just} have introduced universal frameworks to detect attack vectors in DeFi by modeling protocol logic and employing algorithms like Z3 or Bellman-Ford. Large language models (LLMs) also show the potential for anomaly detection in blockchain networks when properly trained ~\cite{gai2023blockchain}. Advanced representations such as Qin et al.~\cite{qin2023towards} execution property graphs can encode runtime details to complement static analysis and improve security for smart contracts in online and forensic settings. In summary, cutting-edge studies are pushing the boundaries of dynamically analyzing, modeling, and securing DeFi ecosystems against exploitation.

\subsection{Large Language Models}
Due to the rise of large models, many scholars have recently conducted in-depth research on them. Ma et al.~\cite{ma2023scope} evaluated the performance of ChatGPT in various subdomains of software engineering. David et al.~\cite{david2023need} studied the detection ability of large models such as \textit{Claude}~\cite{claude} and \textit{GPT}~\cite{gpt-4} for actual attacks on smart contracts. Unlike this, our research focused on several common vulnerabilities in the field of smart contracts, evaluated the effectiveness and limitations of ChatGPT in detecting smart contract vulnerabilities, analyzed the reasons why ChatGPT produced false positives, and compared it with other tools. Li et al.~\cite{li2023test} studied the limitations of large language models in generating software failure-induced test cases and proposed a differential prompt method to improve effectiveness. Fan et al.~\cite{fan2023automated} studied whether automatic program repair technology can fix the error solutions generated by LLMs in the \textit{LeetCode} competition. Wang et al.~\cite{wang2023software} analyzed 52 related studies using LLM for software testing from the perspective of software testing. Joshi et al. presented an LLM-based approach to multilingual repair that enables a flipped interaction model for AI-assisted programming in which the user writes code and the assistant suggests fixes for last-mile mistakes~\cite{Joshi_Cambronero}.
	\UseRawInputEncoding
\pdfoutput=1
\section{Conclusion and Future Works}
\label{sec:conclusion}
In this paper, we assessed ChatGPT's capability in detecting smart contract vulnerability. We formulated three research questions (RQs) to explore the efficacy of ChatGPT in detecting smart contract vulnerabilities, compare it with other vulnerability detection tools, and identify limitations in its performance. For RQ1, we devised an optimized prompt to acquire ChatGPT's detection outcomes, consisting of two steps: vulnerability detection and semantic analysis. The findings suggest that, although ChatGPT demonstrated strengths in pinpointing certain vulnerabilities, it also encountered precision-related challenges. We conducted an analysis of the root causes behind ChatGPT's false positives, categorizing them into four distinct groups. For RQ2, we ran other 14 state-of-the-art vulnerability detection tools on the dataset, and compared ChatGPT with them, revealing that although ChatGPT can detect the most types of vulnerability, the detection performance of ChatGPT is not as good as other state-of-the-art tools for 71.4\% (5/7) of vulnerabilities. ChatGPT demonstrates excellent efficiency, but also exhibits a higher proportion of detection failures compared to 71.4\% (10/14) of the tools we evaluated. For RQ3, we evaluated the uncertainty of ChatGPT by submitting the same smart contract to the model for multiple vulnerability detection and recording the frequency of consistent conclusions. The results indicate that approximately 20\% of the conclusions are uncertain. Additionally, we employed ChatGPT to assess smart contracts of varying lengths, aiming to determine the file size threshold that the model can effectively analyze. We conclude that the maximum code size that the gpt-4 model and gpt-3.5 model can detect is around 33KB in file size and 16KB in file size, respectively. These findings underscore its potential as a vulnerability detection tool, but improvements are still needed for a broader adoption in blockchain security.

In the future, we intend to dive deeper into exploring the capabilities of alternative Large Language Models (LLMs) for smart contract vulnerability detection, including but not limited to \textit{Code llama}~\cite{roziere2023code}. Furthermore, our objectives include leveraging more intricate and diverse datasets to further test the capabilities boundary of LLMs within this domain. We also plan to apply techniques such as fine-tuning~\cite{howard2018universal} and retrieval-augmented generation~\cite{lewis2020retrieval} to enhance the performance of ChatGPT in detecting smart contract vulnerabilities. Furthermore, as new versions of LLMs become available, we will reevaluate our research using these updated models.
	\section*{Acknowledgements}
This work is supported by the Open Research Fund of The State Key Laboratory of Blockchain and Data Security, Zhejiang University and the National Natural Science Foundation of China (62332004, 62276279) and Guangdong Basic and Applied Basic Research Foundation (2024B1515020032).
\balance
	\bibliographystyle{ACM-Reference-Format}
	\bibliography{refs}
\end{document}